%Paper: 9112076
%From: POPE@PHYS.TAMU.EDU
%Date: Tue, 31 Dec 1991 12:22:14 CST

%%%%%%%%%%%%%%%%%%%%%%%%%%%%%%%%%%%%%%%%%%%%%%%%%%%%%
%%                                                 %%
%%     Lectures on W algebras and W gravity        %%
%%                                                 %%
%%                C.N. Pope                        %%
%%                                                 %%
%%              USES PLAIN TEX                     %%
%%                                                 %%
%%%%%%%%%%%%%%%%%%%%%%%%%%%%%%%%%%%%%%%%%%%%%%%%%%%%%

%rosetta stone: left curly {, right curly }, left square [, right square ],
%caret », underscore _, tilde ^, percent %, dollar $, hash #, backslash \,
%forward slash /, ampersand &, exclamation !, left round (, right round )
%left quote `, right quote ', double quotes ", at @
\def\singlespace{\normalbaselines}
\def\oneandahalfspace{\baselineskip=1.15\normalbaselineskip plus 1pt
\lineskip=2pt\lineskiplimit=1pt}

\def\np{\vfill\eject}
\def\nl{\hfil\break}

\def\nofirstpagenoten{\nopagenumbers\footline={\ifnum\pageno>1\tenrm
\hss\folio\hss\fi}}
\def\nofirstpagenotwelve{\nopagenumbers\footline={\ifnum\pageno>1\twelverm
\hss\folio\hss\fi}}
\def\leaderfill{\leaders\hbox to 1em{\hss.\hss}\hfill}
\def\ft#1#2{{\textstyle{{#1}\over{#2}}}}
\def\frac#1/#2{\leavevmode\kern.1em
\raise.5ex\hbox{\the\scriptfont0 #1}\kern-.1em/\kern-.15em
\lower.25ex\hbox{\the\scriptfont0 #2}}
\def\sfrac#1/#2{\leavevmode\kern.1em
\raise.5ex\hbox{\the\scriptscriptfont0 #1}\kern-.1em/\kern-.15em
\lower.25ex\hbox{\the\scriptscriptfont0 #2}}

  %20 point
                   %17 point
  %14 point
 %17 point
 %14 point
 %14 point
 %14 point

\parindent=20pt
\def\narrow{\advance\leftskip by 40pt \advance\rightskip by 40pt}

\def\AB{\bigskip
        \centerline{\bf ABSTRACT}\medskip\narrow}
\def\nonarrower{\advance\leftskip by -40pt\advance\rightskip by -40pt}
\def\AE{\bigskip\nonarrower}

\def\boxit#1{\vbox{\hrule\hbox{\vrule\kern3pt
        \vbox{\kern3pt#1\kern3pt}\kern3pt\vrule}\hrule}}

\def\gtorder{\mathrel{\raise.3ex\hbox{$>$}\mkern-14mu
             \lower0.6ex\hbox{$\sim$}}}
\def\ltorder{\mathrel{\raise.3ex\hbox{$<$}|mkern-14mu
             \lower0.6ex\hbox{\sim$}}}
\def\dalemb#1#2{{\vbox{\hrule height .#2pt
        \hbox{\vrule width.#2pt height#1pt \kern#1pt
                \vrule width.#2pt}
        \hrule height.#2pt}}}

\font\fourteentt=cmtt10 scaled \magstep2
\font\fourteenbf=cmbx12 scaled \magstep1
\font\fourteenrm=cmr12 scaled \magstep1
\font\fourteeni=cmmi12 scaled \magstep1
\font\fourteenss=cmss12 scaled \magstep1
\font\fourteensy=cmsy10 scaled \magstep2
\font\fourteensl=cmsl12 scaled \magstep1
\font\fourteenex=cmex10 scaled \magstep2
\font\fourteenit=cmti12 scaled \magstep1
\font\twelvett=cmtt10 scaled \magstep1 \font\twelvebf=cmbx12
\font\twelverm=cmr12 \font\twelvei=cmmi12
\font\twelvess=cmss12 \font\twelvesy=cmsy10 scaled \magstep1
\font\twelvesl=cmsl12 \font\twelveex=cmex10 scaled \magstep1
\font\twelveit=cmti12
\font\tenss=cmss10
 
 \font\ninebf=cmbx7 scaled \magstep1
\font\ninerm=cmr7 scaled \magstep1 \font\ninei=cmmi7 scaled \magstep1
\font\ninesy=cmsy7 scaled \magstep1 
\font\eightrm=cmr7 scaled 1140 
 
\font\sevenbf=cmbx7 \font\sevenrm=cmr7 \font\seveni=cmmi7
\font\sevensy=cmsy7 

\catcode`@=11
\newskip\ttglue
\newfam\ssfam

\def\fourteenpoint{\def\rm{\fam0\fourteenrm}
\textfont0=\fourteenrm \scriptfont0=\tenrm \scriptscriptfont0=\sevenrm
\textfont1=\fourteeni \scriptfont1=\teni \scriptscriptfont1=\seveni
\textfont2=\fourteensy \scriptfont2=\tensy \scriptscriptfont2=\sevensy
\textfont3=\fourteenex \scriptfont3=\fourteenex \scriptscriptfont3=\fourteenex
\def\it{\fam\itfam\fourteenit} \textfont\itfam=\fourteenit
\def\sl{\fam\slfam\fourteensl} \textfont\slfam=\fourteensl
\def\bf{\fam\bffam\fourteenbf} \textfont\bffam=\fourteenbf
\scriptfont\bffam=\tenbf \scriptscriptfont\bffam=\sevenbf
\def\tt{\fam\ttfam\fourteentt} \textfont\ttfam=\fourteentt
\def\ss{\fam\ssfam\fourteenss} \textfont\ssfam=\fourteenss
\tt \ttglue=.5em plus .25em minus .15em
\normalbaselineskip=16pt
\abovedisplayskip=16pt plus 4pt minus 12pt
\belowdisplayskip=16pt plus 4pt minus 12pt
\abovedisplayshortskip=0pt plus 4pt
\belowdisplayshortskip=9pt plus 4pt minus 6pt
\parskip=5pt plus 1.5pt
\setbox\strutbox=\hbox{\vrule height12pt depth5pt width0pt}
\let\sc=\tenrm
\let\big=\fourteenbig \normalbaselines\rm}
\def\fourteenbig#1{{\hbox{$\left#1\vbox to12pt{}\right.\n@space$}}}

\def\twelvepoint{\def\rm{\fam0\twelverm}
\textfont0=\twelverm \scriptfont0=\ninerm \scriptscriptfont0=\sevenrm
\textfont1=\twelvei \scriptfont1=\ninei \scriptscriptfont1=\seveni
\textfont2=\twelvesy \scriptfont2=\ninesy \scriptscriptfont2=\sevensy
\textfont3=\twelveex \scriptfont3=\twelveex \scriptscriptfont3=\twelveex
\def\it{\fam\itfam\twelveit} \textfont\itfam=\twelveit
\def\sl{\fam\slfam\twelvesl} \textfont\slfam=\twelvesl
\def\bf{\fam\bffam\twelvebf} \textfont\bffam=\twelvebf
\scriptfont\bffam=\ninebf \scriptscriptfont\bffam=\sevenbf
\def\tt{\fam\ttfam\twelvett} \textfont\ttfam=\twelvett
\def\ss{\fam\ssfam\twelvess} \textfont\ssfam=\twelvess
\tt \ttglue=.5em plus .25em minus .15em
\normalbaselineskip=14pt
\abovedisplayskip=14pt plus 3pt minus 10pt
\belowdisplayskip=14pt plus 3pt minus 10pt
\abovedisplayshortskip=0pt plus 3pt
\belowdisplayshortskip=8pt plus 3pt minus 5pt
\parskip=3pt plus 1.5pt
\setbox\strutbox=\hbox{\vrule height10pt depth4pt width0pt}
\let\sc=\ninerm
\let\big=\twelvebig \normalbaselines\rm}
\def\twelvebig#1{{\hbox{$\left#1\vbox to10pt{}\right.\n@space$}}}

\def\tenpoint{\def\rm{\fam0\tenrm}
\textfont0=\tenrm \scriptfont0=\sevenrm \scriptscriptfont0=\fiverm
\textfont1=\teni \scriptfont1=\seveni \scriptscriptfont1=\fivei
\textfont2=\tensy \scriptfont2=\sevensy \scriptscriptfont2=\fivesy
\textfont3=\tenex \scriptfont3=\tenex \scriptscriptfont3=\tenex
\def\it{\fam\itfam\tenit} \textfont\itfam=\tenit
\def\sl{\fam\slfam\tensl} \textfont\slfam=\tensl
\def\bf{\fam\bffam\tenbf} \textfont\bffam=\tenbf
\scriptfont\bffam=\sevenbf \scriptscriptfont\bffam=\fivebf
\def\tt{\fam\ttfam\tentt} \textfont\ttfam=\tentt
\def\ss{\fam\ssfam\tenss} \textfont\ssfam=\tenss
\tt \ttglue=.5em plus .25em minus .15em
\normalbaselineskip=12pt
\abovedisplayskip=12pt plus 3pt minus 9pt
\belowdisplayskip=12pt plus 3pt minus 9pt
\abovedisplayshortskip=0pt plus 3pt
\belowdisplayshortskip=7pt plus 3pt minus 4pt
\parskip=0.0pt plus 1.0pt
\setbox\strutbox=\hbox{\vrule height8.5pt depth3.5pt width0pt}
\let\sc=\eightrm
\let\big=\tenbig \normalbaselines\rm}
\def\tenbig#1{{\hbox{$\left#1\vbox to8.5pt{}\right.\n@space$}}}
\let\rawfootnote=\footnote \def\footnote#1#2{{\rm\parskip=0pt\rawfootnote{#1}
{#2\hfill\vrule height 0pt depth 6pt width 0pt}}}

\def\tenfoot{\tenpoint\hskip-\parindent\hskip-.1cm}

\twelvepoint
\def\sbullet{\raise.2em\hbox{$\scriptscriptstyle\bullet$}}
\nofirstpagenotwelve
%\magnification=\magstep1%
\hsize=16.5 truecm
%\vsize=23.0 truecm
\baselineskip 15pt
%\parskip 0pt
%\font\ftf=cmr8

\def\ft#1#2{{\textstyle{{#1}\over{#2}}}}

\def\lagr{{\cal L}}
\def\tJ{\widetilde J}
\def\th{\tilde h}
\def\tB{\widetilde B}

\def\bG{\overline G}
\def\tW{\widetilde W}
\def\del{\partial}
\def\FF#1#2#3#4#5{\,\sb{#1}F\sb{\!#2}\!\left[\,
{{#3}\atop{#4}}\,;{#5}\,\right]}
\def\tV{\widetilde V}
\def\tg{\tilde g}
\def\tc{\tilde c}
\def\pb{\overline\psi}
\def\phib{\overline\phi}

\def\frac#1#2{{{#1}\over{#2}}}
\def\noss{\noalign{\smallskip}}
\def\sb#1{\lower.4ex\hbox{${}_{#1}$}}

\def\.{\,\,,\,\,}

\def\crampest{\medmuskip = 1mu plus 1mu minus 1mu}
\def\uncramp{\medmuskip = 4mu plus 2mu minus 4mu}

\def\uncramp{\medmuskip = 4mu plus 2mu minus 4mu}

\oneandahalfspace
\rightline{CTP TAMU--103/91}
\rightline{December 1991}

\vskip 2truecm
\centerline{\bf Lectures on $W$ algebras and $W$
gravity\footnote{$^\dagger$}{\tenfoot Lectures given at the Trieste Summer
School in High-Energy Physics, August 1991.}}
 \vskip 1.5truecm \centerline{C.N.
Pope\footnote{$^\star$}{\tenfoot Supported in part by the U.S. Department of
Energy, under grant DE-FG05-91ER40633.}}
\bigskip
\centerline{\it Center for Theoretical Physics, Texas A\&M University,}
\centerline{\it College Station, TX 77843--4242, USA.}

\vskip 1.5truecm
\AB\singlespace
      We give a review of the extended conformal algebras, known as $W$
algebras, which contain currents of spins higher than 2 in addition to the
energy-momentum tensor.  These include the non-linear $W_N$ algebras; the
linear $W_\infty$ and $W_{1+\infty}$ algebras; and their super-extensions.
We discuss their applications to the construction of $W$-gravity and
$W$-string theories.

\AE\oneandahalfspace
\np

\noindent
{\bf Foreword}
\bigskip

     This paper is a review of developments in the area of $W$ algebras, and
their application to the construction of $W$-gravity theories in two
dimensions.  It is based on a series of four lectures presented at the {\it
Summer School in High-energy Physics and Cosmology}, at the ICTP, Trieste, in
the summer of 1991.  The material presented here is largely the result of
collaborations with a number of people, principally Larry Romans and Shawn
Shen.  Special thanks are also due to my other collaborators: Eric Bergshoeff,
Paul Howe, Keke Li, Hong Lu, Ergin Sezgin, Kelly Stelle, Xujing Wang, Kaiwen Xu
and Kajia Yuan.

     The organisation of these lectures is as follows:
\bigskip

\item{1.} Introduction to $W$ algebras.

\item{2.} The $W_3$ and $W_N$ algebras.

\item{3.} The $W_\infty$ and $W_{1+\infty}$ algebras.

\item{4.} Realisations of $W_\infty$ and $W_{1+\infty}$.

\item{5.} BRST operators for $W$ algebras.

\item{6.} Classical and Quantum $W$ gravity.

\bigskip
\noindent{\bf 1. Introduction to $W$ algebras; Virasoro $\rightarrow
W_N\rightarrow W_\infty$}
\bigskip

     The Virasoro algebra has played a very important r\^ole in physics in
the last few years, because it is the essential underlying worldsheet symmetry
of string theory, and of two-dimensional gravity.  In the language of Laurent
modes, it takes the form
$$
[L_m,L_n]=(m-n)L_{m+n} +{c\over 12}m(m^2-1)\delta_{m+n,0},\eqno(1.1)
$$
where the indices $m,n,\ldots$ range over all the integers.  The second term
on the right-hand side is the central term in the algebra, which plays a
crucial r\^ole in the quantum theory.

     The generators $L_m$ can be viewed as the coefficients in a Laurent
expansion of the holomorphic energy-momentum tensor in two dimensions:
$$
T(z)\equiv T_{zz}(z)=\sum_{n=-\infty}^\infty L_n\, z^{-n-2}.\eqno(1.2)
$$
Equivalently, we may write
$$
L_n={1\over 2\pi i}\oint dz \ T(z)  z^{n+1},\eqno(1.3)
$$
where the contour is taken to enclose the origin.  The commutation relations
(1.1) can then be expressed in terms of the operator-product expansion
$$
T(z)T(w)\sim {\partial T(w)\over z-w} +{2 T(w)\over (z-w)^2} +{c/2\over
(z-w)^4},\eqno(1.4)
$$
where the tilde indicates that terms that are non-singular as $z$ approaches
$w$ are omitted.  The procedure for going back and forth between the language
of commutation relations of modes, and operator-product expansions of
currents, is straightforward, and is nicely explained in many reviews on
conformal field theory (see, for example, [1,2,3]).

     The current $T(z)$ is the generator of holomorphic coordinate
transformations in two dimensions.  Acting on fields $\Phi_{\mu_1\cdots
\mu_s}$, it transforms the holomorphic component $\Phi\equiv \Phi_{z\cdots z}$
according to the rule
$$
\delta\Phi(w)={1\over 2\pi i}\oint dz k(z) T(z)\  \Phi(w),\eqno(1.5)
$$
where $k(z)$ is the infinitesimal parameter of the holomorphic
reparametrisation;
$$
\delta z=k(z).\eqno(1.6)
$$
Assuming that $\Phi_{\mu_1\cdots\mu_s}$ transforms as an $s$-index tensor
under general coordinate transformations,
$$
\delta\Phi_{\mu_1\cdots\mu_s}=k^\nu\partial_\nu \Phi_{\mu_1\cdots \mu_s}
+(\partial_{\mu_1}k^\nu) \Phi_{\nu\mu_2\cdots\mu_s} +\cdots +(\partial_{\mu_s}
k^\nu)\Phi_{\mu_1\cdots \mu_{s-1}\nu},\eqno(1.7)
$$
then we deduce that under holomorphic transformations, $\Phi$ transforms as
$$
\delta\Phi=k\partial\Phi +s\, \partial k\, \Phi.\eqno(1.8)
$$
Thus from (1.5) we find that the operator-product expansion of $T$ with
$\Phi$ must be
$$
T(z)\Phi(w)\sim {\partial\Phi(w)\over z-w} +{s\Phi(w)\over(z-w)^2}.\eqno(1.9)
$$

     A field whose OPE with $T(z)$ is of the form (1.9) is called a primary
field of conformal weight (or spin) $s$ (we are considering purely
holomorphic fields here).  Note that the OPE (1.4) for the energy-momentum
tensor $T$ itself has operator terms of the form (1.9), with $s=2$.  However,
the presence of the central term means that $T(z)$ is not a primary field.  The
fact that the $1/(z-w)^n$ terms for $n=1,2,3$ have the standard form (1.9)
means that $T(z)$ is what is called a {\it quasi-primary} field of conformal
weight 2.  The significance of these particular terms in the OPE is that they
correspond to the commutators of $L_{-1}$, $L_0$ and $L_1$ with the Laurent
modes of the field $\Phi$.  For a field of spin $s$, these Laurent modes are
defined by a generalisation of (1.3):
$$
\Phi_n={1\over 2\pi i}\oint dz\ \Phi(z) z^{n+s-1}.\eqno(1.10)
$$
Conversely, the field $\Phi(z)$ is given in terms of its Laurent modes by
$$
\Phi(z)=\sum_{n=-\infty}^\infty \Phi_n\, z^{-n-s}.\eqno(1.11)
$$
As one can see from (1.1), the Virasoro modes $L_{-1}$, $L_0$ and $L_1$ satisfy
the subalgebra
$$
[L_{-1},L_1]=-2L_0,\qquad [L_0,L_{\pm1}]=\mp L_{\pm1}.\eqno(1.12)
$$
This algebra can be recognised as being precisely that of $SL(2,R)$.  It is
called an anomaly-free subalgebra of the Virasoro algebra, since the
(anomalous) central terms in the algebra (1.1) vanish for these particular
commutators.

     We have seen, then, that the Virasoro algebra can be viewed as the
algebra of the (quasi)-primary spin-2 current $T(z)$.  It is natural now to
think about the possibility of extending this algebra by the inclusion of
currents of higher (plus, possibly, lower) spins.  In fact extensions with
only lower spins added have been studied for a long time.  Examples include
the $N=1$ super-Virasoro algebra, with an additional spin-$\ft32$ current
$G(z)$, and the $N=2$ super-Virasoro algebra, with two extra spin-$\ft32$
currents $G(z)$ and $\bG(z)$ and a spin-1 current $J(z)$.  The non-vanishing
OPEs of this $N=2$ algebra are:
$$
\eqalign{
T(z)T(w)&\sim {\partial T\over z-w}+{2T\over (z-w)^2}+{c/2\over(z-w)^4},\cr
T(z)J(w)&\sim{\partial J\over z-w}+{J\over (z-w)^2},\qquad
J(z)J(w)\sim{c/3\over (z-w)^2},\cr T(z)G(w)&\sim {\partial G\over
z-w}+{3/2G\over(z-w)^2},\qquad  T(z)\bG(w)\sim {\partial \bG\over
z-w}+{3/2\bG\over(z-w)^2},\cr J(z)G(w)&\sim {G\over z-w},\qquad J(z)\bG(w)\sim
{-\bG\over z-w},\cr
G(z)\bG(w)&\sim {2T+\partial J\over z-w}+{2J\over (z-w)^2} +{2c/3\over
(z-w)^3}.\cr}\eqno(1.13)
$$
Note that here, and for the rest of this paper, it is to be understood that
unspecified arguments of currents appearing on the right-hand side of OPEs
are taken to be $w$. We see from (1.13) that the currents $G$ and $\bG$
indeed satisfy the primary-field condition (1.9) with $s=\ft32$, and $J$ is a
primary field with spin 1.

     The simplest higher-spin extension of the Virasoro algebra was found by
Zamolodchikov in 1985 [4].  Called $W_3$, it comprises two currents; the
energy-momentum tensor $T(z)$, and a spin-3 primary current $W(z)$.
Subsequently, higher-spin generalisations known as $W_N$ algebras were
constructed [5,6].  These comprise the energy-momentum tensor $T(z)$, together
with higher-spin primary currents of each spin $3\le s\le N$.  We may think
of the Virasoro algebra itself as being the $W_2$ algebra in this
classification.

     A characteristic feature of all the $W_N$ algebras with $N>2$ (and
$<\infty$) is that they are non-linear.  The reason for this is that the
OPE of currents with spins $s$ and $s'$ produces terms that, at leading
order, have spin $s+s'-2$.  If one attempts to build an algebra with
fundamental currents with spins $2\le s\le N$, it follows that that there
will be terms appearing from OPEs that have spins exceeding $N$.  In the
$W_N$ algebras, these are interpreted as composite fields, built from
products of the fundamental currents themselves.  Of course $W_2$ is an
exception, since $2+2-2=2$.

     Another exception to the general rule occurs if $N=\infty$, since now
there is no spin that can be produced in an OPE that is too high to be
represented by the fundamental currents of the algebra;
$\infty+\infty-2=\infty$.  So one might expect, and indeed one finds, that it
should be possible to construct a linear $W_\infty$ algebra, generated by
currents with spins $2\le s\le \infty$ [7,8].  Apart from its linearity, it
otherwise has many features in common with the finite-$N$ $W_N$ algebras; in
particular, in both cases there are non-trivial central terms in the OPE of
any pair of equal-spin currents.  Another related linear algebra with an
infinite number of currents is called $W_{1+\infty}$ [9].  This algebra
contains currents of all spins in the range $1\le s\le \infty$.

     These various $W$ algebras, their super-extensions, and applications to
$W$ gravity, will form the subject of the rest of these lectures.  We begin,
in the next section, by looking in more detail at the finite-$N$ $W_N$
algebras, and in particular, $W_3$.

\bigskip
\noindent{\bf 2. The $W_3$ and $W_N$ algebras}
\bigskip

\noindent{\it 2.1 The $W_3$ algebra}\bigskip

     The $W_3$ algebra comprises two currents; the spin-2 energy-momentum
tensor $T(z)$, and the spin-3 primary current $W(z)$.  The operator products
of these currents are [4]:
$$
\eqalignno{
T(z) T(w)& \sim {\partial T(w)
\over z-w} +{2 T\over (z-w)^2} +{c/2\over
(z-w)^4}&(2.1a) \cr
T(z) W(w)&\sim {\partial W
\over z-w} +{3 W\over (z-w)^2}&(2.1b)\cr
W(z)W(w)& \sim {16\over 22+5c}\Big({\partial \Lambda\over
z-w}+{2\Lambda\over (z-w)^2}\Big)\cr
&\ \ +\ft1{15}\Big({\partial^3 T\over z-w} +\ft92{\partial^2 T\over (z-w)^2}
+15{\partial T\over (z-w)^3} +30{T\over
(z-w)^4}\Big)\cr
&\ \  +{c/3 \over(z-w)^6}&(2.1c)\cr}
$$
The first of these is just the usual Virasoro algebra, and the second
reflects the fact that $W(z)$ is a primary spin-3 current.

   The third operator product requires more explanation.  To leading order,
the OPE of two spin-3 currents produces spin 4, and the quantity $\Lambda$ in
(2.1$c$) denotes the composite spin-4 field.  It is given in terms of a
quadratic product of two energy-momentum tensors, by
$$
\Lambda\equiv (TT)-\ft3{10}\partial^2 T.\eqno(2.2)
$$
The quadratic product $(TT)$ also requires explanation.  Since $T$ is an
operator, it follows that the product of two $T$'s at the same point will be
singular.  To define $(TT)$ we must therefore split the points, and take a
limit after first subtracting the singular terms:
$$
(TT)(w)\equiv \lim_{z\to w}\Big(T(z)T(w)-\hbox{ singular
terms}\Big).\eqno(2.3)
$$
A particularly elegant way to do this was introduced in [10]:  we may adopt the
definition, for an arbitrary pair of operators $A$ and $B$;
$$
(AB)(w)\equiv {1\over 2\pi i}\oint dz {A(z)B(w)\over z-w}.\eqno(2.4)
$$
It is easy to see that this indeed corresponds to a regularisation of the
point-split expression, of the form (2.3).  Note that there is no unique way
to define the regularised limit; (2.4) corresponds to a specific, convenient,
choice.  For this choice, an appropriate way to define multiple
normal-ordered products is:
$$
(ABC\cdots DE)\equiv (A(B(C(\cdots(DE))))).\eqno(2.5)
$$

     It should be emphasised that the normal ordering of the product $(TT)$
is being performed with respect to the modes of $T$ itself, and {\it not}
with respect to the modes of whatever matter fields might be involved in a
specific realisation of $T$.   Because the definition of $(TT)$ is ``self
contained,'' it follows that, despite the non-linearities, we can view the
$W_3$ algebra (2.1$a$-$c$) as being closed on the generators $T$ and $W$.
Although it is not a Lie algebra, it nevertheless satisfies the Jacobi
identities in the usual way, {\it i.e.}
$$
[A,[B,C]]+[B,[C,A]]+[C,[A,B]]=0,\eqno(2.6)
$$
where $A$, $B$ and $C$ are chosen in any combination from $\{L_m\}$ and
$\{W_m\}$, the Laurent modes of $T$ and $W$.

\bigskip
\noindent{\it 2.2 The $W_N$ algebras, and the Miura transformation}
\bigskip

     The $W_3$ algebra described above may be generalised to $W_N$, with one
current of each spin in the interval $2\le s \le N$ [5,6].  In practice, the
explicit structures of the higher-$N$ algebras rapidly become unmanageable.
Already for the $W_4$ algebra, the ``explicit'' expressions are not very
aesthetically attractive [11,12].  For many purpose, however, it is sufficient
to know explicit {\it realisations} of the algebras, even if the details of the
algebras themselves are too complicated for one to wish to see them.  For the
$W_N$ algebras, realisations in terms of $N-1$ free scalars have been given,
by making use of a construction known as the Miura transformation [5]. This
exploits a relation between the $W_N$ algebra and the group $SU(N)$. The way
it works is as follows.  Consider the differential operator $L$, of degree
$N$, given by
$$
L=u_N \del^N +u_{N-2}\del^{N-2}+u_{N-3}\del^{N-3}\cdots +u_1 \del +u_0.
\eqno(2.7)
$$
One now equates $L$ with the factorised differential operator
$$
L=\prod_{i=1}^N\Big(\alpha_0\del +(\vec\mu_i-\vec\mu_{i+1})\cdot \del \vec
\phi\Big),\eqno(2.8)
$$
where $\alpha_0$ is a constant, $\vec\mu_i$ denotes the fundamental weights
of $SU(N)$, and $\vec\phi$ denotes an $(N-1)$-vector of scalar fields.  (The
fundamental weights of $SU(N)$ are defined by $\vec\mu_i\cdot
\vec\alpha_j=\delta_{ij}$, where $\vec\alpha_i$ are the simple roots.)  One
can see by counting derivatives that $u_{N-2}$ has spin 2, $u_{N-3}$ has spin
3, and so on, with $u_0$ having spin $N$.  Essentially, these are the
currents of $W_N$.  In fact, up to a constant scaling, $u_{N-2}$ is precisely
the stress tensor $T$.  For the higher-spin currents, one finds that the
quantities $\tW^{(s)}\equiv u_{N-s}$ themselves are not actually primary
spin-$s$ currents.  To make primary currents $W^{(s)}$, one must add
derivatives and/or products of lower-spin currents.  The details in general
are quite complicated.  For the $W_3$ algebra, the results for the stress
tensor $T$ and the primary spin-3 current $W$ turn out, after redefining the
fields,  to be
\crampest
$$
\eqalignno{
T(z)&=\ft12(\del\phi_1)^2+\ft12(\del\phi_2)^2 +\alpha_1
\del^2\phi_1+\alpha_2 \del^2\phi_2,&(2.9a)\cr
W(z)&={2\sqrt2\over\sqrt{22+5c}}\Big( \ft13(\del\phi_1)^3
-\del\phi_1(\del\phi_2)^2 + \alpha_1 \del\phi_1\del^2\phi_1
-2\alpha_2\del\phi_1\del^2\phi_2 - \alpha_1\del\phi_2\del^2\phi_2\cr
&\qquad\qquad\qquad+\ft13 \alpha_1^2\del^3\phi_1 -\alpha_1\alpha_2
\del^3\phi_2\Big),&(2.9b)\cr}
$$
\uncramp
where $\alpha_1^2=3\alpha_2^2$.  The scalar fields $\phi_i$ satisfy the OPEs
$$
\phi_i(z)\phi_j(w)\sim \delta_{ij}\log(z-w),\eqno(2.10)
$$
from which one can show, after some algebra, that $T$ and $W$ given
by (2.9$a,b$) satisfy the $W_3$ algebra (2.1$a$-$c$) with central charge
$$
c=2-16\alpha_1^2.\eqno(2.11)
$$

    Other bosonic $W$ algebras can also be constructed, which are related to
other Lie groups in an analogous way to the relation between $W_N$ and
$SU(N)$ (see, for example, [13]).  These are qualitatively similar in most
respects to the $W_N$ algebras that we have been discussing, and we shall not
consider them further here.

\vfill\eject
\noindent{\bf 3. The $W_\infty$ and $W_{1+\infty}$ algebras.}
\bigskip

\noindent{\it 3.1 The $W_\infty$ algebra}
\bigskip

     The general structure of the $W_3$ algebra (2.1$a$-$c$), which is also
shared by the higher-$N$ $W_N$ algebras, is that the OPE of currents
$W^{(s)}$ and $W^{(s')}$ of spins $s$ and $s'$ may be cast in the form
$$
W^{(s)}(z)W^{(s')}(w)\sim W^{(s+s'-2)},\ W^{(s+s'-4)},\ W^{(s+s'-6)},\ldots ,
+c_s\delta^{ss'},\eqno(3.1)
$$
where we are being very schematic here, and omitting all the inverse powers
of $(z-w)$.  The sequence of operator terms on the right-hand side descends
in steps of two units of spin, and the final term is the (diagonal) central
term.  We are also using $W^{(s)}$ here to represent either a
fundamental current (if $s\le N$), or a composite current (in general).

     The structure of (3.1) suggests a natural kind of ansatz to try in order
to construct a linear $N\to\infty$ limit of the $W_N$ algebras.  To do this, we
shall, for now, revert to a Laurent-mode description.  Also, for historical
reasons, we shall adopt the notation, of dubious convenience, that a generator
of spin $s$ will be denoted by $V^i$, where $s=i+2$.  Thus we are led to
make the following ansatz:
$$
[V^i_m,V^j_n]=\sum_{\ell\ge 0} g_{2\ell}^{ij}(m,n) V^{i+j-2\ell}_{m+n}
+c_i(m) \delta^{ij}\delta_{m+n,0}.\eqno(3.2)
$$
Here, $g_{2\ell}^{ij}(m,n)$ are the structure constants of the algebra, and
$c_i(m)$ are the central terms.  $V^i_m$ denotes the $m$'th Laurent mode
of the spin-$(i+2)$ current $V^i(z)$.  For the $W_\infty$ algebra, with
spins $2\le s\le\infty$, the indices $i$, $j$ range from 0 to $\infty$.

     The structure constants and central terms can be determined by
demanding that the ansatz (3.2) be consistent with the Jacobi identities.
After considerable brute-force calculation combined with guesswork, one
arrives at the conclusion that [7,8]
$$
c_i(m)=m(m^2-1)(m^2-4)\cdots(m^2-(i+1)^2)c_i,\eqno(3.3)
$$
where $c_i$ are central charges, and the structure constants take the form
$$
g_{\ell}^{ij}(m,n)={1\over 2(\ell+1)!}\phi^{ij}_{\ell}\
N^{ij}_{\ell}(m,n),\eqno(3.4)
$$
where the $N^{ij}_{\ell}(m,n)$ are given by
\crampest
$$
\eqalign{
N_{\ell}^{ij}(m,n)&
=\sum_{k=0}^{\ell+1}(-1)^k{\ell+1\choose k}[i+1+m]_{\ell+1-k}
[i+1-m]_k[j+1+n]_k[j+1-n]_{\ell+1-k},\cr
&=\sum_{k=0}^{\ell+1}(-1)^k{\ell+1\choose
k}(2i+2-\ell)_k[2j+2-k]_{\ell+1-k}[i+1+m]_{\ell+1-k}[j+1+n]_k.\cr} \eqno(3.5)
$$
\uncramp
(The proof of the equivalence of these two expressions is not completely
trivial [8].) In (3.5), $[a]_n$ denotes the descending Pochhammer symbol,
$[a]_n\equiv a(a-1)\cdots (a-n+1)=a!/(a-n)!$, and $a)_n$ denotes the ascending
Pochhammer symbol, $(a)_n\equiv a(a+1)\cdots (a+n-1)= (a+n-1)!/(a-1)!$. The
functions $\phi^{ij}_{\ell}$ are given by
$$
\phi^{ij}_{\ell} =
\FF43{-\ft12\.\ft32\.-\ft{\ell}2-\ft12\.-\ft{\ell}2}{-i-\ft12\.-j-\ft12\.
i+j-\ell+\ft52}1,\eqno(3.6)
$$
where the right-hand side is a
Saalsch\"utzian $_4F_3(1)$ generalised hypergeometric function [8]. In more
down-to-earth language, $\phi^{ij}_\ell$ can be written as
$$
\phi^{ij}_\ell=\sum_{k\ge 0}\frac{(-\ft12)_k(\ft32)_k(-\ft{\ell}2
-\ft12)_k(-\ft{\ell}2)_k}
{k!\,(-i-\ft12)_k(-j-\ft12)_k(i+j-\ell+\ft52)_k}.\eqno(3.7)
$$
The  central charges $c_i$ are given by
$$
c_i={2^{2i-3}i!\, (i+2)!\over(2i+1)!!\, (2i+3)!!}\, c.\eqno(3.8)
$$
These results are unique, modulo trivial redefinitions and rescalings of
generators.  Note that we have given expressions here for $g^{ij}_\ell(m,n)$
for odd as  well as even values of the lower index $\ell$, although only the
even ones appear in (3.2).  This is for later convenience.

     The sequence of generators on the right-hand side of (3.2) descends in
steps of two units of conformal spin.  The lowest-spin generator that can
appear is therefore either 2, if $i+j$ is even, or 3, if $i+j$ is odd.
This termination of the sequence in fact occurs automatically; the
structure constants $g^{ij}_{2\ell}(m,n)$ vanish if $i+j-2\ell\le 0$.
That this should happen is rather non-trivial.  The $N^{ij}_{2\ell}(m,n)$
factor (3.5) in $g^{ij}_{2\ell}(m,n)$ has  ``obvious'' zeros that
are responsible for certain of the necessary zeros in the structure
constants.  The non-triviality lies in the $\phi^{ij}_{2\ell}$ factor given
by (3.6), which supplies the remaining necessary zeros.  The subject of
non-trivial zeros of generalised hypergeometric functions is one that is
much studied by some mathematicians.

     The general results presented above are not very transparent from the
point of view of getting a feel for what the $W_\infty$ algebra really looks
like.  It is instructive, therefore, to look at the first few terms in the
series explicitly.  Thus we have
$$
[V^i_m,V^j_n]=\ft12 \phi^{ij}_0 N^{ij}_0(m,n) V^{i+j}_{m+n} +
\ft1{2\cdot 3!}\phi^{ij}_2 N^{ij}_2(m,n) V^{i+j-2}_{m+n}+
\ft1{2\cdot 5!}\phi^{ij}_4 N^{ij}_4(m,n) V^{i+j-4}_{m+n}+\cdots.\eqno(3.9)
$$
The expressions for $\phi^{ij}_{2\ell}$ and $N^{ij}_{2\ell}(m,n)$ functions
appearing here are:
$$
\eqalign{
\phi^{ij}_0&=1,\cr
\phi^{ij}_2&=1-{9\over (2i+1)(2j+1)(2i+2j+1)},\cr
\phi^{ij}_4&=1-{30\over (2i+1)(2j+1)(2i+2j-3)}\Big(1+{15/2\over (2i-1)(2j-1)
(2i+2j-1)}\Big),\cr}\eqno(3.10)
$$
and
\crampest
$$
\eqalign{
N^{ij}_0(m,n)&=2(j+1)m-2(i+1)n,\cr
N^{ij}_2(m,n)&=4j(j+1)(2j+1)m^3 -12 ij(2j+1)m^2n +12ij(2i+1)mn^2
-4i(i+1)(2i+1)n^3\cr
&\ \ -4j(j+1)(1+3i+3i^2+2j+3ij)m +4i(i+1)(1+3j+3j^2+2i+3ij) n,\cr
N^{ij}_4(m,n)&=\hbox{ 5'{\it th}-order polynomial in $m$ and $n$}.\cr}
\eqno(3.11)
$$
\uncramp
(The details of the last term here are not sufficiently instructive to
justify the space that they would occupy!)

     If one looks at the commutator of the Virasoro modes $L_m\equiv V^0_m$
with $V^j_n$, one will find, from (3.2), that in general the right-hand side
will produce not only $V^j_{m+n}$, from the leading-order term in the sum, but
also lower-spin generators.  This means that in general the generators
$V^j_n$ are not associated with a primary current.  However, it turns out
that if the Laurent index on $L_m$ is restricted to $m=-1,0,1$,
corresponding to the $SL(2,R)$ subalgebra, then the lower-spin terms on the
right-hand side disappear.  This means that the generators $V^j_n$
correspond, in general, to currents that are {\it quasi-primary}.  The
exception is $V^1_n$, associated with the spin-3 current.  This is a genuine
primary current, since for this case it is not possible for there to be any
lower-spin terms on the right-hand side.  Primary currents at all spins could
be defined, but at the price of introducing non-linearities into the algebra.

\bigskip
\noindent{\it 3.2 Contraction to $w_\infty$}
\bigskip

     There is a simple contraction of the $W_\infty$ algebra, which can be
obtained by first rescaling the generators $V^i_m$ by an $i$-dependent power
of a parameter $q$, as follows:
$$
V^i_m\longrightarrow q^{-i} V^i_m.\eqno(3.12)
$$
Thus the commutation relations (3.2) become
$$
[V^i_m,V^j_n]=\sum_{\ell\ge 0} q^{2\ell} g_{2\ell}^{ij}(m,n)
V^{i+j-2\ell}_{m+n} + q^{2i}c_i(m) \delta^{ij}\delta_{m+n,0}.\eqno(3.13)
$$
If we now send the parameter $q\to 0$, then only the highest-spin generator
term on the right-hand side survives ({\it i.e.}\ the $\ell=0$ term), together
with the central term for $i=j=0$.  If we denote the generators for this
contraction of the algebra by $v^i_m$, then from (3.10) and (3.11), we have [7]
$$
[v^i_m,v^j_n]=\Big((j+1)m-(i+1)n\Big) v^{i+j}_{m+n} + {c\over12} m(m^2-1)
\delta^{i,0}\delta^{j,0} \delta_{m+n,0}.\eqno(3.14)
$$
This algebra, now known as $w_\infty$, was discovered some time before the
$W_\infty$ algebra [14].  As we shall see later, it can in some sense be viewed
as the classical limit of the $W_\infty$ algebra.  Because it admits a
central extension only in the Virasoro subalgebra (generated by $L_m=v^0_m$),
it is a somewhat trivial extension of the Virasoro algebra.  From the point
of view of representation theory, the absence of higher-spin central terms
means that the higher-spin states built up from a highest-weight state will
have zero norm, and should be set to zero.  We shall return to a more
detailed discussion of $w_\infty$ and related classical algebras later.

\bigskip
\noindent{\it 3.3 Operator-product expansions for $W_\infty$}
\bigskip

     For many purposes, it is much more convenient to discuss two-dimensional
conformal field theories in terms of fields, and operator-product expansions,
rather than Laurent modes and commutation relations.  This is true also for
the discussion of $W$ algebras, so now we shall look at how to re-express the
content of the $W_\infty$ algebra in the language of conformal field theory.
Following (1.11), we introduce a current $V^i(z)$ for each set of Laurent
modes.  Since $V^i_m$ has (quasi-) conformal spin $(i+2)$, it follows that
the appropriate expansions are
$$
V^i(z)=\sum_{m=-\infty}^\infty V^i_m z^{-m-i-2}.\eqno(3.15)
$$
Conversely, the Laurent modes are given in terms of the currents by
$$
V^i_m={1\over 2\pi i}\oint dz V^i(z) z^{m+i+1}.\eqno(3.16)
$$

     Converting the commutation relations (3.2) for the modes $V^i_m$ into
operator-product expansions for the currents (3.15) is now a purely
mechanical procedure [8].  Essentially, powers of $m$ or $n$ appearing in the
structure constants $g^{ij}_{2\ell}(m,n)$ get replaced by derivatives.  The
result, including the rescaling (3.12) for convenience, is that the OPE of
$V^i(z)$ with $V^j(w)$ is given by
$$
V^i(z)V^j(w)\ \sim\ -\sum_{\ell\ge0} q^{2\ell}f^{ij}_{2\ell}\bigl(\partial_z,
\partial_w\bigr){V^{i+j-2\ell}(w)\over{z-w}}\
-q^{2i}c_i\delta^{ij}(\partial_z)^{2i+3}{1\over{z-w}}.\eqno(3.17)
$$
The quantities $f^{ij}_{2\ell}(m,n)$ are closely related to the structure
constants $g^{ij}_{2\ell}(m,n)$ in (3.2).  In fact the only difference is
that instead of (3.4) we have
$$
f_{\ell}^{ij}(m,n)={1\over 2(\ell+1)!}\phi^{ij}_{\ell}\
M^{ij}_{\ell}(m,n),\eqno(3.18)
$$
where $\phi^{ij}_\ell$ is still given by (3.6), but $M^{ij}_\ell(m,n)$, which
replaces $N^{ij}_\ell(m,n)$, is given by [8]
$$
M_{\ell}^{ij}(m,n)
=\sum_{k=0}^{\ell+1}(-1)^k{\ell+1\choose
k}(2i+2-\ell)_k[2j+2-k]_{\ell+1-k} m^{\ell+1-k}n^k.\eqno(3.19)
$$
This expression is in fact the part of $N^{ij}_\ell(m,n)$ that is of total
degree $\ell+1$ in $m$ and $n$, as may be seen by comparing it with the
second expression in (3.5).  Thus, for example, $M^{ij}_2(m,n)$ is given by
the first line of the expression in (3.11) for $N^{ij}_2(m,n)$.

     It is instructive to look at a few examples of OPEs for the currents of
$W_\infty$.  Setting the parameter $q$ in (3.17) to
$$
q=\ft14\eqno(3.20)
$$
for convenience, we have,
$$
\eqalign{
V^0(z)V^0(w)&\sim {\partial V^0\over z-w}+{2V^0\over (z-w)^2} +{c/2\over
(z-w)^4},\cr
V^0(z)V^1(w)&\sim {\partial V^1\over z-w}+{3V^1\over (z-w)^2},\cr
V^0(z)V^2(w)&\sim {\partial V^2\over z-w}+{4V^2\over (z-w)^2}
+\ft{12}5{V^0\over (z-w)^4},\cr
V^1(z)V^1(w)&\sim {2\partial V^2\over z-w} +{4 V^2\over (z-w)^2}\cr
&\quad +\ft1{10}\Big({\partial^3 V^0\over z-w} +\ft92 {\partial^2 V^0\over
(z-w)^2} +15{\partial V^0\over (z-w)^3} + 30 {V^0\over (z-w)^4}\Big) \cr
&\quad +{c/2\over (z-w)^6}.\cr}\eqno(3.21)
$$
Thus we see that $V^1(z)$ is a primary spin-3 current, whilst $V^2(z)$ (and
indeed all the higher-spin currents) is quasi-primary.  Note that the
relative coefficients of the terms for a given spin on the right-hand side
are given purely by the $M^{ij}_{2\ell}(m,n)$ part of $f^{ij}_{2\ell}(m,n)$.
The relations between these coefficients are in fact governed completely by
covariance under the $SL(2,R)$ subalgebra generated by $L_{-1}$, $L_0$, and
$L_1$; in other words, they reflect the quasi-primary nature of the currents.
The similarity between the OPEs for $V^0$ and $V^1$ in (3.21), and those for
$T$ and $W$ in (2.1$a$-$c$), is quite striking.  The essential difference is
that here $V^2$ is an independent spin-4 current, whereas in the $W_3$
algebra $\Lambda$ is a composite spin-4 current, given in terms of $T$ by
(2.2).

\bigskip
\noindent{\it 3.4 The $SL(\infty,R)$ wedge subalgebra}
\bigskip

     We have already seen that the Virasoro algebra contains a
finite-dimensional subalgebra, namely the $SL(2,R)$ algebra generated by
$L_{-1}$, $L_0$ and $L_1$.  The commutation relations for these generators
are given in (1.12).  Since the Virasoro algebra is a subalgebra of
$W_\infty$, it follows that $SL(2,R)$ is a subalgebra of $W_\infty$ also.  In
fact, $W_\infty$ has a much larger subalgebra, which is a natural extension
of $SL(2,R)$.  One can show from the commutation relations (3.2), with the
definitions of the structure constants given below (3.2), that the subset of
generators $V^i_m$ with Laurent indices in the range
$$
-i-1\le m\le i+1\eqno(3.22)
$$
form a closed algebra [8]. If one plots the ``spin'' index $i$ vertically,
and the Laurent index $m$ horizontally, then the subset of generators
specified by (3.22) fill out a ``wedge,'' with the three Virasoro generators
$L_{-1}$, $L_0$ and $L_1$ sitting at the bottom, followed by the five spin-3
generators $V^1_{-2}$, $V^1_{-1}$, $V^1_0$, $V^1_1$ and $V^1_2$ at the
next-highest level, and so on.  This subset of generators at each level in
fact transforms as the $(2i+3)$-dimensional irreducuble representation of the
$SL(2,R)$ generated by $L_{-1}$, $L_0$ and $L_1$.

     If one takes the canonical embedding of $SL(2,R)$ in $SL(N,R)$ in which
the fundamental ${\bf N}$ of $SL(N,R)$ decomposes irreducibly to the ${\bf N}$
of $SL(2,R)$, then the adjoint of $SL(N,R)$, the $({\bf N^2-1})$, decomposes
under $SL(2,R)$ as
$$
({\bf N^2-1)}\longrightarrow {\bf 3}\ \oplus \ {\bf 5}\ \oplus\ {\bf 7}\
\oplus \cdots\oplus ({\bf 2N-1}).\eqno(3.23)
$$
Thus we see that the set of $SL(2,R)$ representations that we get from the
wedge subalgebra of $W_\infty$ is precisely the set of representations that
arise from the decomposition of $SL(\infty,R)$ under the above canonical
embedding of $SL(2,R)$.  So the wedge subalgebra is $SL(\infty,R)$ [8].

     Unlike finite-dimensional Lie algebras, such as $SL(N,R)$, where all
choices of basis for the generators are equivalent, one can have many {\it
inequivalent} sets of commutation relations for infinte-dimensional algebras
such as $SL(\infty,R)$.  One way to see this is to think of constructing
$SL(\infty,R)$ as an $N\to\infty$ limit of $SL(N,R)$.  Although all basis
choices are equivalent at finite $N$, one may make $N$-dependent
redefinitions to give different bases which, after taking the limit
$N\to\infty$, can no longer be related to one another.  In fact for
$SL(\infty,R)$ there is, for example, a nice one-parameter family of
algebras which are inequivalent, in the sense that no redefinition of
generators enables one to relate the commutation relations of one member of
the family to those for another [15,16,8].  This family of $SL(\infty,R)$
algebras can be constructed by starting from the generators $L_{-1}$, $L_0$
and $L_1$ of $SL(2,R)$, and building the tensor algebra of all products of
arbitrary numbers of $SL(2,R)$ generators, modulo the ideal  $$
x\otimes y-y\otimes x -[x,y]=0. \eqno(3.24)
$$
In other words, the commutation relations of the $L_m$ can be used to
simplify products.  One may now impose a further ideal relation, namely
$$
{\cal I}:\qquad Q-s(s+1)=0,\eqno(3.25)
$$
where $Q\equiv L_0^2-\ft12(L_1 L_{-1} +L_{-1} L_1)$ is the Casimir operator of
$SL(2,R)$, and $s$ is an arbitrary constant.  The commutators of the tensor
operators, modulo these ideals, close on $SL(\infty,R)$.  The constant $s$
parametrises inequivalent $SL(\infty,R)$ algebras.  Without loss of
generality, we may restrict $s$ to the range
$$
s\ge -\ft12.\eqno(3.26)
$$

     One can show that the commutation relations for the $SL(\infty,R)$
generators $X^i_m$,  where $m$ is subject to the ``wedge'' condition (3.22),
are given by
$$
[X^i_m,X^j_n]=\sum_{\ell\ge0} g^{ij}_{2\ell}(m,n;s)X^{i+j-2\ell}_{m+n},
\eqno(3.27)
$$
with
$$
g_{\ell}^{ij}(m,n;s)={1\over 2(\ell+1)!}\phi^{ij}_{\ell}(s)\
N^{ij}_{\ell}(m,n).\eqno(3.28)
$$
Here,
$$
\phi^{ij}_{\ell}(s) =
\FF43{-\ft12-2s\.\ft32+2s\.-\ft{\ell}2-\ft12\.-\ft{\ell}2}{-i-\ft12\.-j-\ft12\.
i+j-\ell+\ft52}1,\eqno(3.29)
$$
and $N^{ij}_\ell(m,n)$ is given by (3.5).  Thus we see that the wedge
subalgebra of $W_\infty$ coincides with the $s=0$ member of the family of
$SL(\infty,R)$ algebras (3.27) [8].

\bigskip
\noindent{\it 3.5 The $W_{1+\infty}$ algebra}
\bigskip

     For all values of $s$, the $SL(\infty,R)$ algebras (3.27) have the
property that the descending sequence of terms on the right-hand side
automatically cuts off at $X^0_{m+n}$, {\it i.e.}\ the superscript on
$X^{i+j-2\ell}_{m+n}$ never becomes negative.  This happens because
$g^{ij}_{2\ell}(m,n;s)$ supplies the necessary zeros.  However, unlike the
case of $W_\infty$, many of the zeros here occur because of the restricted
ranges (3.22) that the Laurent indices can have.  Thus if one were to relax
the restrictions (3.22), and allow the Laurent indices to range over all the
integers, one would find for a generic value of $s$ that the termination
property would be lost, and the sequence of terms in (3.27) would continue
indefinitely.  If one wanted to try to interpret such an algebra as an
algebra of conformal fields, it would correspond to having all conformal
spins from $-\infty$ to $\infty$. This would be undesirable in any physical
application, since the two-point correlation function for fields of negative
conformal spins diverges as the points are separated. It is a very special
property of the $s=0$ case that the termination property persists when the
Laurent indices are extended ``beyond the wedge.''

     Having seen that $W_\infty$ algebra (without central terms) can be viewed
as the extension of (3.27) beyond the wedge, it is natural to ask if there
are any other special values of $s$ for which algebras of only non-negative
conformal spins can be obtained.  It turns out that there is just one other
possibility, corresponding to the case when $s=-\ft12$ [8,9].  In this case,
it turns out that the zeros of the structure constants are such as to terminate
the sequence of generators on the right-hand side at spin 1 (if $i+j$ is odd)
or spin 2 (if $i+j$ is even).  Thus we get an algebra with currents of all
conformal spins $\ge1$.  For this reason, this algebra is referred to as
$W_{1+\infty}$.  Like $W_\infty$, it also admits a central extension.  The
complete result for the $W_{1+\infty}$ algebra is then
$$
[\tV^i_m,\tV^j_n]=\sum_{\ell\ge 0} q^{2\ell}\tg_{2\ell}^{ij}(m,n)
\tV^{i+j-2\ell}_{m+n} +q^{2i}\tc_i(m) \delta^{ij}\delta_{m+n,0},\eqno(3.30)
$$
where
$$
\tg^{ij}_\ell(m,n)\equiv g^{ij}_\ell(m,n,-\ft12)\eqno(3.31)
$$
and
$$
\tc_i(m)=m(m^2-1)(m^2-4)\cdots(m^2-(i+1)^2)\tc_i,\eqno(3.32)
$$
with
$$
\tc_i={2^{2i-2}\big( (i+1)!\big)^2\over (2i+1)!!\, (2i+3)!!}\, c.\eqno(3.33)
$$
We have included the rescaling parameter $q$ in (3.30) for later convenience.
Examples of the first few
$$
\tilde\phi^{ij}_\ell\equiv \phi^{ij}_\ell(-\ft12)\eqno(3.34)
$$
functions are
$$
\eqalign{
\tilde\phi^{ij}_0&=1,\cr
\tilde\phi^{ij}_2&=1+{3\over (2i+1)(2j+1)(2i+2j+1)},\cr
\phi^{ij}_4&=1+{10\over (2i+1)(2j+1)(2i+2j-3)}\Big(1+{27/2\over (2i-1)(2j-1)
(2i+2j-1)}\Big),\cr}\eqno(3.35)
$$
Thus the structure constants are similar in form, but different in detail,
to those for $W_\infty$.

     The $W_{1+\infty}$ algebra may be recast in the language of
operator-product expansions of currents, just as in the $W_\infty$ case.  The
only differences are that now the $\phi^{ij}_\ell$ appearing in (3.18) are
replaced by $\tilde\phi^{ij}_\ell$ given by (3.34) and (3.29), and the $c_i$
given by (3.8) are replaced by the $\tc_i$ given by (3.33).  Thus we have
$$
\tV^i(z)\tV^j(w)\ \sim\ -\sum_{\ell\ge0} q^{2\ell}{\tilde
f}^{ij}_{2\ell}\bigl(\partial_z,
\partial_w\bigr){\tV^{i+j-2\ell}(w)\over{z-w}}\
-q^{2i}\tc_i\delta^{ij}(\partial_z)^{2i+3}{1\over{z-w}},\eqno(3.36)
$$
with
$$
\tilde f_{\ell}^{ij}(m,n)={1\over 2(\ell+1)!}{\tilde\phi}^{ij}_{\ell}\
M^{ij}_{\ell}(m,n),\eqno(3.37)
$$
Examples of OPEs for $W_{1+\infty}$, after setting the parameter $q=\ft14$ for
convenience, are
$$
\eqalign{
\tV^{-1}(z)\tV^{-1}(w)&\sim {c\over (z-w)^2},\cr
\tV^0(z)\tV^0(w)&\sim {\partial \tV^0\over z-w}+{2\tV^0\over (z-w)^2}
+{c/2\over (z-w)^4},\cr
\tV^0(z)\tV^{-1}(w)&\sim {\partial \tV^{-1}\over z-w} +{\tV^{-1}\over
(z-w)^2},\cr
\tV^0(z)\tV^1(w)&\sim {\partial \tV^1\over z-w}+{3\tV^1\over (z-w)^2}
+\ft16 {\tV^{-1}\over (z-w)^4}.\cr }\eqno(3.38)
$$
(Recall that because of the notational convenience that $\tV^i(z)$ has spin
$i+2$, the spin-1 current is denoted by $\tV^{-1}(z)$ here.)

\bigskip
\noindent{\it 3.6 The Lone-star algebra}
\bigskip

     The $W_\infty$ algebra was originally constructed abstractly, in the
sense that a formal antisymmetric Lie bracket $[A,B]=-[B,A]$ was assumed, which
was then required to satisfy the Jacobi identity
$$
[A,[B,C]]+[B,[C,A]]+[C,[A,B]]=0.\eqno(3.39)
$$
These requirements then led to the solution for the structure constants and
central terms of the $W_\infty$ algebra.  Interestingly, there is a way to
realise the abstract bracket as the antisymmetric part of a more fundamental
associative product of generators that we may represent by $\star$.  Thus we
may write
$$
[A,B]\equiv A\star B- B\star A,\eqno(3.40)
$$
where $\star$ satisfies the associativity property
$$
A\star(B\star C)=(A\star B)\star C.\eqno(3.41)
$$
In terms of this realisation, the Jacobi identity (3.39) is now trivially
satisfied, by virtue of the associativity of the $\star$ product.  The
``lone-star'' product of generators $V^i_m$ and $V^j_n$ of $W_\infty$ turns
out to be given by [9]
$$
V^i_m\star V^j_n=\ft12 \sum_{\ell\ge -1}q^\ell g^{ij}_\ell(m,n)
V^{i+j-\ell}_{m+n}, \eqno(3.42)
$$
where the suffix $\ell$ on $g^{ij}_\ell(m,n)$ now takes odd-integer as well as
even-integer values.  It is for this reason that the definition of
$g^{ij}_\ell(m,n)$ was originally given in (3.4-3.7) for all integers.  The
terms in (3.42) with even $\ell$ are antisymmetric under the simultaneous
interchange $i\leftrightarrow j$ and $m\leftrightarrow n$, whilst the terms
with odd $\ell$ are symmetric.  Showing that (3.42) defines an associative
product is non-trivial.   Some examples of low-lying lone-star products for
$W_\infty$, with $q$ chosen to be $\ft14$,  are:
$$
\eqalign{
V^0_m\star V^0_n&=V^1_{m+n} +\ft12 (m-n)V^0_{m+n},\cr
V^0_m\star V^1_n&= V^2_{m+n} +\ft12(2m-n) V^1_{m+n} +\ft1{20}(6m^2 -3mn
+n^2-4) V^0_{m+n},\cr
V^1_m\star V^0_n&=V^2_{m+n} +\ft12(m-2n) V^1_{m+n} +\ft1{20} (m^2-3m n+6n^2
-4) V^0_{m+n}.\cr}\eqno(3.43)
$$

     A lone-star product may be defined for $W_{1+\infty}$ also [9].  It is
just like (3.42), only now $g^{ij}_\ell(m,n)$ is replaced by
$\tg^{ij}_\ell(m,n)$, given by (3.31).  Low-lying examples for $W_{1+\infty}$
are:
$$
\eqalign{
\tV^{-1}_m\star \tV^{-1}_n&=\tV^{-1}_{m+n},\cr
\tV^{-1}_m\star \tV^0_n&= \tV^0_{m+n} +\ft12 m \tV^{-1}_{m+n},\cr
\tV^0_m\star \tV^{-1}_n&= \tV^0_{m+n} -\ft12 n \tV^{-1}_{m+n},\cr
\tV^0_m\star \tV^0_n&=\tV^1_{m+n} +\ft12 (m-n) \tV^0_{m+n} +\ft1{12} (m^2-m n
+n^2 -1) \tV^{-1}_{m+n}.\cr}\eqno(3.44)
$$
Note that in these, and indeed in all lone-star products for $W_{1+\infty}$,
the generator $V^{-1}_0$ acts like the identity operator in the algebra.

\bigskip
\noindent{\bf 4. Realisations of $W_\infty$ and $W_{1+\infty}$}
\bigskip

\noindent{\it 4.1 Differential-operator realisations at $c=0$}
\bigskip

   Before turning to the consideration of quantum realisations of $W_\infty$
and $W_{1+\infty}$ in terms of scalar or spinor fields, it is interesting to
look at purely {\it classical} realisations of the (centreless) algebras.
They can in fact be realised in terms of enveloping algebras of simple (in
the non-technical sense) infinite-dimensional algebras.  Let us define, for
convenience,
$$
\eqalign{
j_n&\equiv \tV^{-1}_n,\cr
L_n&\equiv \tV^0_n.\cr}\eqno(4.1)
$$
Then one can show, from the general form of the lone-star product, that
$$
\eqalignno{
j_m\star j_n&=j_{m+n},&(4.2)\cr
L_0\star \tV^i_m &=\tV^{i+1}_m -\ft12 m \tV^i_m -{(i+1)^2 \big[
(i+1)^2-m^2\big] \over 4\big[4(i+1)^2-1\big] } \tV^{i-1}_m.&(4.3)\cr}
$$
This latter equation can be turned around and viewed as a recursive
definition of $\tV^{i+1}_m$ in terms of $\tV^i_m$ and $\tV^{i-1}_m$ [17].
Starting from a humble $U(1)$ Kac-Moody algebra, with generators $j_m$,
together with a derivation operator $d$:
$$
\eqalignno{
[j_m,j_n]&=0,&(4.4)\cr
[d,j_m]&=-m\, j_m,&(4.5)\cr}
$$
we may first construct the Virasoro generators $L_m$ as
$$
L_m=(d+\ft12 m)j_m,\eqno(4.6)
$$
where, motivated by (4.2),  we impose the ideal relation ${\cal I}$ defined by
$$
{\cal I}:\qquad j_m j_n -j_{m+n}=0.\eqno(4.7)
$$
Then, motivated by (4.3), we define
$$
\tV^i_m\equiv (d+\ft12 m)\tV^{i-1} +{i^2(i^2-m^2)\over 4(4i^2-1)}
\tV^{i-2}_m.\eqno(4.8)
$$
This implies that [17]
$$
\tV^i_m=P_i(d;m)j_m,\eqno(4.9)
$$
where the first few $P_i(d;m)$ are given by
$$
\eqalign{
P_{-1}&=1,\cr
P_0&=d+\ft12 m,\cr
P_1&= d^2+m d +\ft1{12}(2m^2+1),\cr
P_2&= d^3 +\ft32 m d^2 +\ft1{20} (12m^2 +7) +\ft1{40}
(2m^3+7m),\cr
P_3&=d^4+2m d^3+\ft1{14}(18m^2+13)d^2+\ft1{14}(4m^3+13m)d+\ft1{560}
(8m^4+100m^2+27).\cr}\eqno(4.10)
$$
It follows from (4.3) that the $\tV^i_m$ defined by (4.8) will generate the
(centreless) $W_{1+\infty}$ algebra [17].

     The above construction shows that the $W_{1+\infty}$ algebra, without
central extension, can be realised as the Lie algebra of the enveloping
algebra of a centreless $U(1)$ Kac-Moody algebra with a derivation operator
$d$ [17].  One specific realisation for $j_m$ and $d$ is provided by taking
$$
\eqalign{
j_m&=e^{im\theta},\cr
d&=i{\partial\over\partial\theta},\cr}\eqno(4.11)
$$
where $\theta$ is a coordinate on a circle of unit radius.  One can easily
verify that these expressions for $j_m$ and $d$ satisfy the commutation
relations (4.4) and (4.5), and that the ideal-relation (4.7) is automatically
implied.  Thus we see that the centreless $W_{1+\infty}$ algebra can be viewed
as the algebra of all polynomials in $e^{im\theta}$ and
$i\partial/\partial\theta$ on a circle.  In other words, {\it the centreless
$W_{1+\infty}$ algebra is isomorphic to the algebra of all smooth
differential operators, of arbitrary degree, on the circle} [17].  The Virasoro
subalgebra is generated by all smooth differential operators of degree 1 on
the circle.

     A very similar discussion can be given for the $W_\infty$ algebra.  The
analogues of (4.2) and (4.3) now are [17]
$$
\eqalignno{
L_m\star L_n&=V^1_{m+n}+\ft12 (m-n)L_{m+n},&(4.12)\cr
L_0\star V^i_m&=V^{i+1}_m-{m\over 2} V^i_m
-{i(i+2)[(i+1)^2-m^2]\over 4[
4(i+1)^2-1]}V^{i-1}_m.&(4.13)\cr}
$$
Motivated by these, we can now  construct $W_\infty$ as the Lie algebra of
the enveloping algebra of the Virasoro algebra, modulo the ideal ${\cal J}$
suggested by (4.12):
$$
{\cal J}:\qquad L_m L_n -(L_0+m) L_{m+n}=0.\eqno(4.14)
$$
Again, one can solve recursively for the $V^i_m$, to give
$$
V^i_m=Q_i(L_0;m) L_m.\eqno(4.15)
$$
The first few examples are
$$
\eqalign{
Q_0&=1,\cr
Q_1&=L_0+\ft{m}2,\cr
Q_2&=L_0^2+m L_0+\ft15(m^2+1),\cr
Q_3&=L_0^3+\ft{3m}2 L_0^2+\ft1{14} (9m^2+10)L_0
+\ft1{14}(m^3+5m),\cr
Q_4&=L_0^4+2m L_0^3+\ft13(4m^2+5)L_0^2+\ft13(m^3+5m)L_0
+\ft1{42}(m^4+15m^2+8).\cr}\eqno(4.16)
$$
In this case, we can realise the Virasoro algebra, and the ideal relation
(4.14), by taking
$$
L_m=i e^{im\theta} {\partial\over\partial\theta}.\eqno(4.17)
$$
Thus we see that the $W_\infty$ algebra without central extension is
isomorphic to the algebra of all smooth differential operators of strictly
positive degree on a circle [17].

     The above discussions show that $W_\infty$ must be a subalgebra of
$W_{1+\infty}$, since the former is the algebra of all smooth differential
operators of positive degree on the circle, whilst the latter is the algebra
of all smooth differential operators of non-negative degree [17].  The
embedding of $W_\infty$ in $W_{1+\infty}$ is not completely obvious in the
usual bases that we have been using.  For example, one can see from the last
OPE in (3.36) that one cannot simply set the spin-1 current to zero in
$W_{1+\infty}$ in that basis, since it is generated on the right-hand side of
OPEs of higher-spin currents.  By comparing the above representations for
$W_{1+\infty}$ and $W_\infty$, one can show that $W_\infty$ may in fact be
obtained as a truncation of $W_{1+\infty}$, after an appropriate redefinition
of the generators.  If we denote the generators of $W_{1+\infty}$ by
$\tV^i_m$, and those of $W_\infty$ by $V^i_m$ as usual, then these
redefinitions take the form
$$
V^i_m=\sum_{j=-1}^i a_{ij}(m)
\tV^j_m,\eqno(4.18)
$$
where the constants $a_{ij}(m)$ are polynomials in $m$ of
degree $(i-j)$, and $i$ takes the values $0,1,2,\ldots$. For example, we have
$$
\eqalign{
V^0_m&=\tV^0_m+\ft{m}2 \tV^{-1}_m-\ft
c8\delta_{m,0},\cr
V^1_m&=\tV^1_m+\ft{m}2
\tV^0_m+\ft1{12}(m^2-1)\tV^{-1}_m,\cr
V^2_m&=\tV^2_m+\ft
m2\tV^1_m+\ft1{20}(2m^2-3)\tV^0_m
+\ft1{120}(2m^3+7m)\tV^{-1}_m-\ft{3c}{320}\delta_{m,0}.\cr}\eqno(4.19)
$$
The generators $V^i_m$, $i\ge 0$, together with $\tV^{-1}_m$, yield
the $W_{1+\infty}$ algebra in a new basis. After making the redefinitions
(4.18), the conformal-spin-1 generator $\tV^{-1}_m$ can be consistently
truncated from the $W_{1+\infty}$ algebra. The remaining generators yield
$W_\infty$. Note that, for convenience, we have included constant shifts in the
definitions of $V^i_0$, which ensure that the ``trivial''
lower-order parts of the central terms have their canonical form.  The
resulting central charge $c$ of the $W_\infty$ algebra is related
to the central charge $\tc$ of $W_{1+\infty}$ by [17]
$$
c=-2{\tc}.\eqno(4.20)
$$
As is well known, in order to have unitary representations of algebras of
this kind (for example Kac-Moody and Virasoro algebras), one requires that the
central charge be positive.  Thus (4.20) suggests that unitary representations
of $W_{1+\infty}$ may not straightforwardly decompose  into unitary
representations of $W_\infty$.

\bigskip
\noindent{\it 4.2 Fermionic realisations of $W_{1+\infty}$ and $W_\infty$}
\bigskip

     Consider a free complex fermion $\psi$, with OPE
$$
\pb(z)\psi(w)\sim {1\over z-w}.\eqno(4.21)
$$
This may be used to give a realisation of the Virasoro algebra, by
defining the energy-momentum tensor $T$ to be
$$
T(z)=\ft12 \del\pb\, \psi-\ft12 \pb\, \del\psi.\eqno(4.22)
$$
One may easily verify from (4.21) that this satisfies the Virasoro algebra
(1.4) with central charge $c=1$.  One may also build a spin-1 current that is
bilinear in $\pb$ and $\psi$, namely $J(z)=\pb\psi$.  It is straightforward
to verify that this satisfies the OPEs
$$
\eqalign{
J(z)J(w)&\sim {1\over (z-w)^2},\cr
T(z)J(w)&\sim {\del J\over z-w} +{J\over (z-w)^2}.\cr}\eqno(4.23)
$$
The currents $T$ and $J$ give a closed algebra; in fact precisely the bosonic
subalgebra of the $N=2$ superconformal algebra (1.13).

     One may go on to contruct higher-spin currents that are bilinear in
$\pb$ and $\psi$, by adding more derivatives.  Now, one finds that the OPE
algebra of the currents will not close on any finite set of higher-spin
currents (at least if we insist that the algebra be linear).  By including
currents for {\it all} higher spins however, closure can be achieved [18].  As
one might perhaps expect, the algebra that one obtains is precisely
$W_{1+\infty}$ [19,20].  Depending on how one chooses to define the higher-spin
currents, one obtains $W_{1+\infty}$ in different bases.  No matter what
(non-degenerate) choice is made, however, it will always be $W_{1+\infty}$
in some basis.  The reason for this is that there is in fact only one
independent current that can be built at each spin, in the sense that there
is only one new current at a given spin that is not simply a linear
combination of derivatives of lower-spin currents.  For example, at spin 2
there are two independent currents that can be written down; $\del\pb\, \psi$,
and $\pb\, \del\psi$.  One linear combination of these is nothing but $\del J$,
where $J=\pb\psi$; the new, independent, current is $T$, defined by (4.22).
At spin 3 there are three independent currents that can be built; one
combination of these is $\del^2 J$, another is $\del T$, and the remaining
independent combination corresponds to the spin-3 current of $W_{1+\infty}$.
All the higher-spin currents work in a similar fashion.

     A particularly convenient choice of bases for the currents is one for
which they are all quasi-primary with respect to the energy-momentum tensor.
It turns out that after imposing this additional requirement, one is left with
a 1-parameter family of bases [17,21].  The currents take the form:
$$
V^i(z)=\sum_{j=0}^{i+1} \alpha_j(i;a) \del^j \pb \del^{i+1-j}\psi,
\eqno(4.24)
$$
where the coefficients are given by [21]
$$
\alpha_j(i;a)={i+1\choose j}{(i+2a+2-j)_j (2a-i-1)_{i+1-j} \over (i+2)_{i+1}}
\eqno(4.25)
$$
and $a$ parametrises the choice of basis.  The first few currents take the
form
$$
\eqalign{
V^{-1}&=\pb\, \psi,\cr
\noss
V^{0}&=(a+\ft12)\partial\pb\,\psi +(a-\ft12) \pb\,
\partial\psi,\cr
\noss
V^1&=\ft13(a+\ft12)(a+1)\partial^2\pb\, \psi +\ft23
(a+1)(a-1)\partial\pb\, \partial\psi +\ft13
(a-\ft12)(a-1)\pb\, \partial^2\psi.\cr
\noss
V^2&=\ft1{60}(2a+3)(2a+1)(a+1)\partial^3\pb\, \psi +
\ft1{20}(2a+3)(2a-3)(a+1)\partial^2\pb\, \partial\psi
\cr
&+\ft1{20}(2a+3)(2a-3)(a-1) \partial\pb \, \partial^2
\psi + \ft1{60}(2a-3)(2a-1)(a-1)\pb\, \partial^3 \psi.
\cr}\eqno(4.26)
$$

     For any value of the parameter $a$, the currents (4.24) generate the
$W_{1+\infty}$ algebra.  The standard basis, in which the OPEs take the form
(3.36) (with $q=\ft14$), corresponds to $a=0$.  Then the first few currents
take the form
$$
\eqalign{
V^{-1}&=\pb\, \psi,\cr
\noss
V^{0}&=\ft12\partial\pb\,\psi -\ft12 \pb\,
\partial\psi,\cr
\noss
V^1&=\ft16\partial^2\pb\, \psi -\ft23 \partial\pb\, \partial\psi
+\ft16 \pb\, \partial^2\psi.\cr
\noss
V^2&=\ft1{20}\partial^3\pb\, \psi -
\ft9{20}\partial^2\pb\, \partial\psi +\ft9{20}\partial\pb \,
\partial^2 \psi -\ft1{20}\pb\, \partial^3 \psi.
\cr}\eqno(4.27)
$$
This gives a realisation of $W_{1+\infty}$ with central charge $c=1$.  In
general however, at generic values of $a$, the central terms will not be
diagonal in spin; only at $a=0$, and $a=\pm\ft12$, will they be diagonal.
The two cases $a=\pm\ft12$ are equivalent modulo a conjugation automorphism
of the algebra.  These correspond to the redefinitions of generators (4.18)
for which $W_\infty$ can be embedded in $W_{1+\infty}$.  Thus, as one can
readily check, at $a=\pm\ft12$ we find that the central charge is given by
$c=-2$, in accordance with the general result (4.20).  At $a=\ft12$, the
first few currents are:
$$
\eqalign{
V^{-1}&=\pb\, \psi,\cr
V^0&=\partial{\overline\psi}\, \psi,\cr
V^1&=\ft12 \partial^2{\overline\psi}\, \psi-\ft12\partial{\overline\psi}
\,\partial
\psi,\cr
V^2&=\ft15\partial^3{\overline\psi} \, \psi-\ft35 \partial^2
{\overline\psi}\,
\partial\psi +\ft15\partial{\overline\psi}\, \partial^2\psi.\cr}\eqno(4.28)
$$
In this basis, one can consistently omit the spin-1 current $V^{-1}(z)$;
the remaining currents $V^i(z)$ with $i\ge 0$ generate the $W_\infty$
algebra in the standard basis, as in (3.17) (with $q=\ft14$).

\bigskip
\noindent{\it 4.3 Bosonic realisations of $W_\infty$ and $W_{1+\infty}$}
\bigskip

     In section, we shall consider two bosonic realisations of the algebras.
The first of these is a realisation of $W_\infty$, in terms of currents built
from bilinears in a free complex scalar field $\phi$, with OPE
$$
\phib(z) \phi(w)\sim \log(z-w).\eqno(4.29)
$$
It is easy to see that the spin-2 current $\del\phib\del\phi$ generates
the Virasoro algebra with central charge $c=2$.  As in the fermionic
realisation discussed previously, there is just one independent higher-spin
current that can be built at each spin, by acting with additional derivatives
on the two fields in the bilinear products.  For any non-degenerate choice of
higher-derivative currents, therefore, their OPEs will close on the same
algebra.  This turns out to be $W{_\infty}$ [22].  The choice
$$
V^i(z)={2^{-i-1}(i+2)!\over (i+1)(2i+1)!!}\sum_{k=0}^i(-)^k {i+1 \choose k}
{i+1\choose k+1} \del^{i-k+1}\phib \, \del^{k+1}\phi \eqno(4.30)
$$
generates the $W_\infty$ algebra in the standard basis (3.17) [22], with the
scaling parameter $q$ taking the value $\ft14$.  The first few $W_\infty$
currents in this realisation take the form
$$
\eqalign{
V^0&=\del\phib\, \del\phi,\cr
V^1&=\ft12\del\phib\, \del^2\phi -\ft12\del^2\phib\, \del\phi,\cr
V^2&=\ft15 \del\phib\, \del^3\phi -\ft35 \del^2\phib\,\del^2\phi +\ft15
\del^3\phib\, \del\phi.\cr}\eqno(4.31)
$$

   Finally in this section on bosonic realisations, let us consider the
bosonisation of the one-parameter family of fermionic realisations of
$W_{1+\infty}$ given by (4.24).  The complex fermion $\psi$, and its
conjugate $\pb$, may be written in terms of a free real scalar $\varphi$, as
$$
\psi=:e^\varphi:\qquad \pb=:e^{-\varphi}:,\eqno(4.32)
$$
where $\varphi$ satisfies the OPE
$$
\varphi(z)\varphi(w)\sim\log(z-w).\eqno(4.33)
$$
As shown in [23], the fermion bilinear $\partial^i \pb\,\partial^j
\psi$ can be expressed as
$$
:\partial^i \pb(z)\, \partial^j \psi(z): =\sum_{k=i+1}^{i+j+1}{1\over k}
(-)^{k-1-i}{\ell\choose k-1-i} \partial^{i+j-k+1} P^{(k)}(z),\eqno(4.34)
$$
where $P^{(k)}(z)$ is given by
$$
P^{(k)}(z)=:e^{-\varphi(z)}\partial^k e^{\varphi(z)}:.\eqno(4.35)
$$
Using this, the currents of $W_{1+\infty}$ in the basis (4.24) can then
be written in bosonised form.  The first few bosonised currents are given by
$$
\eqalign{
V^{-1}&=\partial\varphi,\cr
\noss
V^0&=\ft12(\partial\varphi)^2 +a \partial^2\varphi,\cr
\noss
V^1&=\ft13(\partial\varphi)^3 +a\partial\varphi\, \partial^2\varphi
+\ft13a^2\partial^3\varphi,\cr
\noss
V^2&=\ft14(\partial\varphi)^4+a (\partial\varphi)^2\partial^2\varphi
+\ft1{20}(8a^2-3)(\partial^2\varphi)^2
+\ft1{10}(4a^2+1)\partial\varphi\, \partial^3\varphi
+\ft1{60}a(4a^2+1)\partial^4\varphi.\cr}\eqno(4.36)
$$

\bigskip
\noindent{\it 4.4 The super-$W_\infty$ algebra}
\bigskip

      By combining the complex-fermion realisation (4.24) of $W_{1+\infty}$,
and the complex-boson realisation (4.30) of $W_\infty$, we may easily
construct an $N=2$ superalgebra whose bosonic sector is $W_{1+\infty}\times
W_\infty$ [19].  The free Lagrangian $L=\bar\del\phib\, \del\phi
+\pb\bar\del \psi$ has a rigid $N=2$ supersymmetry, and so it is guaranteed
that if we form all possible bilinear currents from the fields $\phi$ and
$\psi$, then they will generate an $N=2$ superalgebra.  Thus in addition to
the bosonic currents (4.24) (with the parameter $a$ taken to be zero for
convenience) and (4.30), we may build fermionic currents $G^i(z)$ and
$\bG^i(z)$, of the form
$$
\eqalign{
G^i(z)&=\sum_{k=0}^i \gamma_k(i)\del^{i-k+1}\phib\, \del^k\psi,\cr
\bG^i(z)&=\sum_{k=0}^i \gamma_k(i)\del^{i-k+1}\phi\, \del^k\pb.\cr}
\eqno(4.37)
$$
A convenient choice for the coefficients $\gamma_k(i)$ is [19]
$$
\gamma_k(i)={2^{-i}(i+1)!(-)^k\over (2i+1)!!} {i\choose k} {i+1\choose
k}.\eqno(4.38)
$$
One can now check that the $W_{1+\infty}$ currents $\tV^i$ ($i\ge-1$), the
$W_\infty$ currents $V^i$ ($i\ge 0$), and the fermionic currents $G^i$ and
$\bG^i$ ($i\ge0 $) close on a superalgebra, which we may call the $N=2$
super-$W_\infty$ algebra [19].  We shall not present details here; they are
quite complicated, and may be found in [19].  Structurally, the form of the
algebra is
$$
\eqalign{
V V&\sim V,\qquad \tV\tV\sim \tV,\qquad V\tV\sim 0,\cr
V G&\sim G,\qquad \tV G\sim G,\qquad V \bG\sim \bG,\qquad \tV\bG \sim \bG,\cr
G\bG &\sim V\oplus \tV,\qquad G G\sim 0,\qquad \bG \,\bG\sim 0.\cr} \eqno(4.39)
$$

\bigskip
\noindent{\bf 5. The BRST operator and anomaly freedom}
\bigskip

\noindent{\it 5.1 The BRST operator for the Virasoro algebra}
\bigskip

     The construction of the BRST operator for the Virasoro algebra is well
known [24].  One introduces anticommuting fields $c(z)$ of conformal spin
$-1$ and $b(z)$ of conformal spin 2, which are the ghost and antighost for
the spin-2 current $T(z)$.  They satisfy the OPE
$$
b(z)c(w)\sim {1\over z-w}.\eqno(5.1)
$$
The BRST operator $Q$ is then defined in terms of the BRST current $j(z)$:
$$
Q={1\over 2\pi i}\oint dz\, j(z),\eqno(5.2)
$$
where $J(z)$ is given by
$$
j(z)=T c +c\del c\, b.\eqno(5.3)
$$
The form of the cubic ghost term in (5.3) is dictated by the structure
constants of the Virasoro algebra, and can be deduced from the general
expression
$$
Q=X^a c_a -\ft12 f^{ab}{}_c c_a c_b b^c\eqno(5.4)
$$
for the BRST operator for a Lie group $G$ with generators $X^a$ satisfying
the algebra $[X^a,X^b]=f^{ab}{}_c X^c$.  The BRST operator $Q$ should be
nilpotent;
$$
Q^2=0.\eqno(5.5)
$$
For a finite-dimensional Lie algebra, this is an automatic consequence of the
definition (5.4), by virtue of the fact that the algebra satisfies the Jacobi
identity.  For an infinite-dimensional algebra such as the Virasoro algebra,
however, there can be a non-trivial central term in the algebra, and this
gives rise to the possibility of a BRST anomaly.

     Nilpotence of $Q$ defined by (5.2) is equivalent to the OPE of
$j(z)j(w)$ being a total derivative.  For the BRST current (5.3) for the
Virasoro algebra, one finds that, up to a total derivative,
$$
j(z)j(w)\sim \Big( \ft12 c-13\Big){c(z)c(w)\over (z-w)^4} ,\eqno(5.6)
$$
where $c$ is the central charge in the Virasoro algebra generated by $T(z)$.
(There will be no confusion between the central charge $c$ and the ghost
field $c$.)  Thus there is a BRST anomaly unless the central charge satisfies
$$
c=26.\eqno(5.7)
$$
This is the celebrated result that implies that the bosonic string is
anomaly free in 26 spacetime dimensions.  Another way of expressing the
result is that the ghost current $T_{\rm gh}(z)$, defined by $T_{\rm
gh}=T_{\rm tot}-T$ with $T_{\rm tot}(z)=\{Q,b(z)\}$, yields a realisation of
the Virasoro algebra with central charge $c_{\rm gh}=-26$, and this must be
cancelled by an equal and opposite central charge from the matter realisation
$T$.  The ghost current is given by
$$
T_{\rm gh}=-2b\del c -\del b\, c.\eqno(5.8)
$$
The BRST operator may be rewritten as
$$
Q={1\over 2\pi i}\oint dz \Big(T(z)+\ft12 T_{\rm gh}(z)\Big) c(z).\eqno(5.9)
$$

     If one writes the BRST operator in the language of Laurent modes, then it
is straightforward to establish also that the intercept is $L_0=1$, {\it
i.e.}\  that nilpotency of $Q$ requires not only $c=26$, but also that
$Q=(L_0-1)c_0 +\sum_{m\ne0} L_m c_{-m} +\sum_{m<n} (m-n):b_{m+n} c_{-m}
c_{-n}$.

\vfill\eject

\noindent{\it 5.2 The BRST operators for $W_\infty$, $W_{1+\infty}$ and super
$W_\infty$}
\bigskip

     For the case of $W$ algebras, with higher-spin currents, one must
introduce ghost-antighost pairs for each such current.  We shall postpone for
now the discussion of how to construct the BRST operator for non-linear
algebras such as $W_N$, and consider the more straightforward linear
$W_\infty$-type algebras.  We shall describe the example of $W_\infty$ in
some detail; the discussion goes through {\it mutatis mutandis} for
$W_{1+\infty}$ and super $W_\infty$.

     For the $W_\infty$ algebra, given by (3.17), there will be a ghost
field $c_i(z)$ of spin $s=-i-1$, and an antighost field $b_i(z)$ of spin
$s=i+2$ for each current $V^i(z)$.  From the general expression (5.4), one
can show that the BRST current for the $W_\infty$ algebra will then be
given by [25]
$$
j(z)=V^i c_i -\sum_{\ell\ge0} f^{ij}_{2\ell}(\del_{c_i},\del_{c_j})c_i c_j
b_{i+j-2\ell},\eqno(5.10)
$$
where summation over $i$ and $j$ is understood.  In this expression,
$\del_{c_i}$ denotes $\del/\del z$ acting {\it only} on the ghost field
$c_i$.  The BRST operator may be written as
$$
Q={1\over 2\pi i}\oint dz \Big(V^i(z) +\ft12 V^i_{\rm gh}(z)\Big) c_i(z),
\eqno(5.11)
$$
where the ghost currents $V^i_{\rm gh}(z)$ are given by
$$
V^i_{\rm gh}(z)=\sum_{\ell\ge0} f^{ij}_{2\ell}(-\del,\del_{c_j}) c_j
\,b_{i+j-2\ell},\eqno(5.12)
$$
where the derivative $\del$ denotes $\del/\del z$ acting on both $c_j$ and
$b_{i+j-2\ell}$, whilst $\del_{c_j}$ denotes $\del/\del z$ acting only on
$c_j$.

    The ghost currents (5.12) should furnish a realisation of the $W_\infty$
algebra, with a certain central charge $c_{\rm gh}$.  Nilpotence of the BRST
operator $Q$ will then be achieved if the matter realisation $V^i(z)$ has
central charge $c=-c_{\rm gh}$.   The general structure of the OPE algebra for
the currents $V^i_{\rm gh}(z)$ turns out to be:
$$
V^i_{\rm gh}(z)V^j_{\rm gh}(w)\ \sim\ -\sum_{\ell\ge 0}
f^{ij}_{2\ell}\bigl(\partial_z, \partial_w\bigr){V^{i+j-2\ell}_{\rm gh}(w)
\over{z-w}} +{C_{ij}\over (z-w)^{i+j+4}},\eqno(5.13)
$$
so the operator terms on the right-hand side are precisely those of the
$W_\infty$ algebra.  However a difficulty that arises in this
case is that the expression (5.12) for the ghost currents involves sums over
infinitely-many ghost fields, and consequently the central terms in the OPE
algebra of the ghost currents will have divergent coefficients $C_{ij}$.  A
natural procedure, therefore, is to try to regularise the divergent sums.
The hope would be that the regularised coefficients $C_{ij}$ would turn out
to have precisely the right form for $W_\infty$ with some central charge
$c_{\rm gh}$, {\it i.e.}
$$
C_{ij}={2^{2i-3} i!\, (2i+3)!\, (i+2)!\over (2i+1)!!\, (2i+3)!!}\, c_{\rm gh}
\delta_{ij},\eqno(5.14)
$$
(see (3.8) and (3.17), with $q=1$).  In fact, one could adopt the principle
that the criterion for a ``good'' regularisation scheme is that it should
yield precisely (5.14); after all, we know that the only possible structure for
central terms that are compatible with the Jacobi identities is precisely
that given in (5.14).

     As is often the case, the naivest guess for how to regularise divergent
quantities seems to be the right one.  The forms of the divergent sums for
$C_{ij}$ at arbitrary $i$ and $j$ are extremely complicated, involving
contributions from the higher-order terms in (5.12); we shall return to these
in a moment.  Consider first, however, the $C_{00}$ term in (5.13).  It is
easy to see that this receives contributions only from the leading-order
terms in (5.12); {\it i.e.}
$$
V^i_{\rm gh}(z)=(i+j+2)\del c_j\, b_{i+j} +(j+1) c_j\, \del b_{i+j}
+\cdots ,\eqno(5.15)
$$
where a summation over $j$ is understood.  It is now easy to check that the
ghosts $c_j$ and $b_j$ for the spin $s=(j+2)$ current $V^j$ contribute a term
$$
C_{00}(s)=-(6s^2-6s+1)\eqno(5.16)
$$
to $C_{00}$, yielding the divergent result
$$
C_{00}=-\sum_{s\ge2}(6s^2-6s+1).\eqno(5.17)
$$
A naive guess for regularising this would be to interpret the sums in terms of
analytically-continued Riemman zeta functions.  Thus, we may write
$$
C_{00}=-\sum_{n\ge0}\Big[ 6(n+\ft32)^2-\ft12\Big],\eqno(5.18)
$$
and then formally reinterpret $C_{00}$ as [26,25]
$$
C_{00}=6\zeta(-2,\ft32)-\ft12 \zeta(0,\ft32),\eqno(5.19)
$$
where $\zeta(s,a)$ is the generalised Riemann zeta function, defined as the
analytic continuation from $\Re(s)>1$ of the function
$$
\zeta(s,a)=\sum_{n\ge 0} (n+a)^{-s}.\eqno(5.20)
$$
{}From (5.19) we find $C_{00}=1$, and hence using (5.14) we get [26,25]
$$
c_{\rm gh}=2.\eqno(5.21)
$$
Of course if this were the only calculation of the regularised value for
$c_{\rm gh}$ it would be hard to take it seriously.  One could arrive at any
desired answer, by choosing the regularisation scheme appropriately.  To give
just one example, the second term in the summand of (5.18) could equally well
be reinterpreted as $-\ft12\zeta(0,a)$ for {\it any} value of $a$, thus giving
$$
c_{\rm gh}=\ft12(2a+1).\eqno(5.22)
$$
In fact, however, the requirement that {\it all} of the $C_{ij}$ coefficients
should regularise consistently, to give (5.14), puts very stringent
conditions on the scheme that one can use.  Whilst one could, of course,
still arrange to get any desired value for $c_{\rm gh}$, the scheme that did
this would look increasingly contrived as one proceeded to higher and higher
spin anomalies.  There really seems to be only one regularised value for
$c_{\rm gh}$ that can be arrived at in a ``natural'' and uncontrived way, and
that is precisely the value given in (5.21) [25].  In [25], a regularisation
scheme is proposed, and checked up to the spin-18 level (by computer!), that
yields this result in a consistent way.

     If one repeats the analysis for the $W_{1+\infty}$ algebra, the effect
on the calculation of $C_{00}$ is to include the contribution of the $s=1$
term in (5.17).  This implies that $C_{00}$, and hence $c_{\rm gh}$, is zero
for $W_{1+\infty}$.  Again, the self-consistent scheme proposed in [25] gives
this same result for a wide range of examples that have been checked.  There
is in fact a further check on the whole scheme that can now be performed,
since in the $W_{1+\infty}$ algebra there will also be an anomaly term in
(5.13) involving $C_{-1,-1}$.  This spin-1 anomaly in fact receives no
contributions from ghosts, and so it is necessarily zero.  Thus without any
regularisation at all we know that the answer must be $c_{\rm gh}=0$ for
$W_{1+\infty}$, and so it is a reassuring check on the validity of the
regularisation scheme that it reproduces the already-correct result.

    A similar analysis has also been carried out for the super-$W_\infty$
algebra [25].  In this case, it turns out that the total ghost central charge
in the Virasoro sector is $c_{\rm gh}=3$.  In summary, therefore, the values of
the central charges in the {\it matter} sector that are needed to achieve
anomaly freedom are:
$$
\eqalign{
W_\infty:&\qquad c=-2,\cr
W_{1+\infty}:&\qquad c=0,\cr
\hbox{super-$W_\infty$}:&\qquad c=-3.\cr}\eqno(5.23)
$$
By writing the BRST operators in terms of the Laurent modes of the currents,
one can also determine the intercepts.  As for the central terms, these
require regularisation.  It turns out that all intercepts regularise to zero
for all the algebras $W_\infty$, $W_{1+\infty}$ and super $W_\infty$ [25].
All these results may be recast into the form of (regularised) Ricci-curvature
conditions for certain K\"ahler coset manifolds of the form $W/H$, where $W$
is the group corresponding to the particular $W$ algebra in question, and $H$
is a subgroup [27].  This description generalises that given for the Virasoro
algebra in [28].

\bigskip
\noindent
{\it 5.3 The BRST operator for $W_3$}
\bigskip

     Although the $W_3$ algebra is non-linear, the construction of the BRST
operator follows rather similar lines to that for a linear algebra.  In
particular, we still have an expression of the form (5.11) for $Q$, namely
$$
Q=\oint dz \Big( \big(T+\ft12 T_{\rm gh}\big) c +\big( W+\ft12 W_{\rm gh}
\big) \gamma \Big),\eqno(5.24)
$$
where $(b,c)$ are the ghosts for $T$, and $(\beta,\gamma)$ are the ghosts for
$W$.  The matter currents $T$ and $W$ generate the $W_3$ algebra (2.1$a$-$c$).
The non-linearity of the algebra reflects itself in the fact that the ghost
current $W_{\rm gh}$ now involves the matter current $T$ as well as the ghost
fields [29,30].  Since the ghost current $T_{\rm gh}$ still has the standard
form, it follows that the ghost central charge in the spin-2 sector is still
given by the analogue of (5.17), leading to the requirement that the matter
central charge $c$ be given by [29]
$$
c=26+74=100.\eqno(5.25)
$$
The ghost currents $T_{\rm gh}$ and $W_{\rm gh}$ are then given by [29,30]
$$
\eqalignno{
T_{\rm gh}&=-2b\,\partial c-\partial
b\, c-3\beta\, \partial\gamma-2\partial\beta\, \gamma&(5.26a)\cr
W_{\rm gh}&=-\partial\beta\,
c-3\beta\, \partial c-\ft8{261}\big[\partial(b\, \gamma\,  T )+b\,
\partial\gamma \, T\big]\cr &\ \ +\ft{25}{6\cdot261}\hbar
\Big(2\gamma\, \partial^3b+9\partial\gamma\, \partial^2b
+15\partial^2\gamma\,\partial b+10\partial^3\gamma\,  b\Big).&(5.26b)\cr}
$$
Note that the spin-3 ghost current $W_{\rm gh}$ involves the spin-2 matter
current $T$.  This looks intuitively reasonable; one can view the non-linear
terms on the right-hand side of (2.1$c$) as being like a linear
algebra but with $T$-dependent structure ``constants.''  These structure
constants then appear in the construction of the ghost currents.  Note also
that the ghost currents need not, and indeed do not, satisfy the $W_3$ algebra
[29,30].  It is shown in these references that, provided the matter central
charge is given by (5.25), the BRST operator (5.24) is indeed nilpotent.

     In [29,30], the BRST operator is actually constructed in terms of the
Laurent modes $L_m$ and $W_m$ of $T(z)$ and $W(z)$.  Nilpotency of $Q$ not
only implies that $c$ should equal 100, but it also determines the values of
the intercepts for $L_0$ and $W_0$.  These turn out to be $L_0=4$, and
$W_0=0$ [29,30].

\vfill\eject

\noindent{\bf 6. Classical and quantum $W$ gravity}
\bigskip

\noindent{\it 6.1 $W_\infty$ gravity}
\bigskip

          The classical theory of $w_\infty$ gravity was constructed in [31].
In its simplest form, one can consider a {\it chiral} gauging of $w_\infty$;
{\it i.e.}\ one gauges just one copy of the algebra, in, say, the left-moving
sector of the two-dimensional theory.  We shall discuss non-chiral gaugings in
more detail later.  For now we just remark that, thanks to an ingenious trick
introduced in [32], involving the use of auxiliary fields, the treatment of the
non-chiral case can be essentially reduced to two independent copies of the
chiral case.

     As our starting point, let us consider the free action $S=1/\pi \int
d^2z L$ for a single scalar field in two dimensions, where $L$ is given by
$$
L=\ft12 \bar\partial \varphi\, \partial\varphi. \eqno(6.1)
$$
Here, we use coordinates $z=x^-$ and $\bar z=x^+$ on the (Euclidean-signature)
worldsheet. This action is invariant under the semi-rigid spin-$s$
transformations
$$
\delta\varphi=\sum_{s\ge2} k_s (\partial\varphi)^{s-1},\eqno(6.2)
$$
where the parameters $k_s$ are functions of $z$, but not $\bar z$.
The spin-$s$ transformation, with parameter $k_s$, is generated by the
current $v^i(z)$, with $i=s-2$:
$$
v^i(z)={1\over i+2}
(\partial\varphi)^{i+2}.\eqno(6.3)
$$
At the classical level, these currents generate the $w_\infty$ algebra.  In
operator-product language, this means that they close on the $w_\infty$
algebra at the level of {\it single} contractions.  This is given in
Laurent-mode language in (3.14).  To keep
track of the orders it is useful to introduce Planck's constant, so that the
OPE of the field $\varphi$  is
$$
\partial\varphi(z)\partial\varphi(w)\sim {\hbar\over(z-w)^2}.\eqno(6.4)
$$
The OPEs of the currents are then given by
$$
\hbar^{-1} v^i(z) v^j(w)\sim (i+j+2){v^{i+j}(w)\over (z-w)^2}+ (i+1)
{\partial v^{i+j}(w)\over z-w} +O(\hbar).\eqno(6.5)
$$
The $\hbar$-independent terms on the right-hand side, corresponding to single
contractions in the OPE, are precisely those for the $w_\infty$ algebra.

     The classical semi-rigid $w_\infty$ symmetry (6.2) of (6.1) can be gauged
by introducing a gauge field $A_i$ for each current $v^i$.  Thus we find that
the Lagrangian
$$
L=
\ft12 \bar\partial \varphi\, \partial\varphi-\sum_{i\ge0} A_i v^i\eqno(6.6)
$$
is invariant under local $w_\infty$ transformations [31], where the gauge
fields are assigned the transformation rules:
$$
\delta A_i=\bar\partial k_i -\sum_{j=0}^i\Big( (j+1)A_j \partial
k_{i-j} -(i-j+1)k_{i-j}\partial A_j\Big).\eqno(6.7)
$$

     When one is presented with a classical theory it is natural, when
considering its quantisation, to begin by contemplating what might go wrong.
In the case of a gauge theory, with classical local symmetries, the
obvious danger is that these might become anomalous at the quantum level.
Indeed, in the case of two-dimensional gravity, the gauge theory of the
Virasoro algebra, we know that anomaly freedom requires that the matter fields
should generate the Virasoro algebra with central charge $c=26$, in order to
cancel the central-charge contribution of $-26$ from the ghosts for the gauge
fixing of the spin-2 gauge field (the metric).  We can certainly expect to meet
analogous anomalies in the higher-spin generalisations that we are considering
here.  In fact, potentially-worse things could also happen:

     Commonly, as for example in the case of the Virasoro algebra and
two-dimensional gravity, one has matter currents that are quadratic in matter
fields.  These, by construction, generate the gauge algebra at the classical
(single-contraction) level.  Upon quantisation, higher numbers of contractions
must also be taken into account, corresponding to Feynman diagrams with closed
loops.  If the currents are at most quadratic in matter fields, then the
``worst case'' is to have two contractions between a pair of currents.  This
corresponds therefore to a pure $c$-number term in the OPE of currents; in
fact, the central term in the Virasoro algebra.  It is these terms that in
fact save the critical two-dimensional gravity theory from anomalies, by
cancelling against anomalies from the ghost sector.

     Things are potentially worse in the case of $w_\infty$ gravity because now
the currents (6.3) involve arbitrarily-high powers of the matter field
$\varphi$.  Thus at the quantum level, one might get matter-dependent
anomalies, associated with diagrams corresponding to multiple contractions of
the currents (6.3) that still have some matter fields left uncontracted.  Of
course the question of whether a theory is actually anomalous is really a
cohomological one, in the sense that the crucial question is whether or not it
is possible to introduce finite local counterterms, and $\hbar$-dependent
renormalisations of the classical transformation rules, in such a way as to
remove the apparently-anomalous contributions of the kind we have been
considering.  Only if such an attempt fails can the theory be said to be
anomalous.

     In general, the process of quantising a theory, and introducing
counterterms and renormalisations of the transformation rules order by order in
$\hbar$ to remove potential anomalies, can be a complicated and tedious one.
Fortunately, in our two-dimensional example the process of removing potential
matter-dependent anomalies can be accomplished in one fell swoop.  All that we
have to do is to find quantum renormalisations of the classical currents (6.3)
such that at the full quantum level (arbitrary numbers of contractions in the
OPEs) they close on an algebra.  This algebra, whatever it turns out to be,
will be the quantum renormalisation of the original classical $w_\infty$
algebra.  In fact, as we shall see, it is precisely $W_\infty$ [33].

     The problem, then, boils down to the fact that the currents (6.3) do not
close on any algebra at the quantum level.  We must therefore seek
$\hbar$-dependent modifications of them, with a view to achieving quantum
closure.  The most general plausible modifications would consist of
higher-order terms added to (6.3) in which the same number of derivatives (to
give the same spin $s=i+2$) are distributed over smaller numbers of $\varphi$
fields.  From (6.4), we see that $\varphi$ has the dimensions of $\sqrt\hbar$,
and so the modifications will be of the form of power series in $\sqrt\hbar$.
Thus we may try an ansatz of the form
$$
V^i={1\over i+2} (\partial\varphi)^{i+2} + \alpha_i \sqrt\hbar
(\partial\varphi)^i \partial^2\varphi + \beta_i \hbar (\partial
\varphi)^{i-1} \partial^3\varphi +\gamma_i \hbar (\partial\varphi)^{i-2}
(\partial^2\varphi)^2 + O(\hbar^{3/2}),\eqno(6.8)
$$
for constant coefficients $\alpha_i$, $\beta_i$, $\gamma_i,\cdots$ to be
determined.  Requiring quantum closure of the OPE algebra for these currents
will then give conditions on this infinite number of coefficients.  They
will not be determinable uniquely, since one is always free to make
redefinitions of currents of the form $V^i\to V^i +\partial V^{i-1}+\cdots$.
However, if, for convenience and without loss of generality, we demand that the
currents should all be quasi-primary with respect to the energy-momentum tensor
$V^0$, then the result is unique.  The expressions for the first few
renormalised currents (spins 2, 3 and 4) are [33]:
$$
\eqalign{
V^0&=\ft12 (\partial\varphi)^2 +\ft12\sqrt\hbar \partial^2\varphi,\cr
V^1&=\ft13(\partial\varphi)^3+\ft12 \sqrt\hbar \partial\varphi
\partial^2\varphi +\ft1{12}\hbar \partial^3\varphi,\cr
V^2&=\ft14(\partial\varphi)^4 +\ft12\sqrt\hbar
(\partial\varphi)^2\partial^2\varphi -\ft1{20}\hbar (\partial^2\varphi)^2
+\ft15 \hbar \partial\varphi \partial^3\varphi +\ft1{60}\hbar^{3/2}
\partial^4\varphi.\cr}\eqno(6.9)
$$

     These renormalised currents can be recognised as the currents of the
$W_\infty$ algebra.  We saw in section (4.3) that there is a realisation of
$W_{1+\infty}$ in terms of a real scalar $\varphi$, obtained by bosonising
the complex-fermion realisation (4.24).  Setting the parameter $a=\ft12$, so
that the truncation to $W_\infty$ can be performed, and instating the
parameter $\hbar$, we see that the currents (6.9), and their higher-spin
colleagues, are precisely the currents of the bosonisation of the
complex-fermion realision.

     Having obtained renormalised currents that close on an algebra (the
$W_\infty$ algebra) at the full quantum level, we are now guaranteed to
have a quantum theory with no matter-dependent anomalies.  The prescription
for writing down the counterterms and renormalisations of the
transformation rules (6.2) and (6.7) necessary to make this anomaly freedom
manifest is very straightforward.  For the counterterms, we simply replace
the classical currents $v^i$ in the Lagrangian (6.6) by the renormalised
currents $V^i$, of which the first few are given by (6.9).  The terms
independent of $\hbar$ are just the original classical currents, whilst the
$\hbar$-dependent terms are the necessary counterterms.  For the
transformation rules, we use the ones that are generated by the renormalised
currents.  For $\varphi$, this means we have $$
\delta\varphi=\hbar^{-1}\sum_{i\ge0} \oint{dz\over 2\pi i} k_i(z) V^i(z)
\varphi(w).\eqno(6.10)
$$
For $A_i$, we will have
$$
\delta A_i=\bar\partial k_i +\hat\delta A_i,\eqno(6.11)
$$
where $\hat\delta A_i$ is such that
$$
\sum_{i\ge0 }\int\Big( \hat\delta A_i V^i +A_i\delta V^i\Big)=0,\eqno(6.12)
$$
with $\delta V^i$ given by
$$
\delta V^i=\hbar^{-1} \sum_{j\ge0}\oint{dz\over 2\pi i} k_j(z)
V^j(z) V^i(w).\eqno(6.13)
$$
The $\hbar$-independent terms in these transformation rules are precisely
the original classical ones (6.2) and (6.7).  The $\hbar$-dependent terms
are the renormalisations necessary, together with the counterterms, to make
the absence of matter-dependent anomalies manifest.  The fact that the
potential matter-dependent anomalies are actually removable by this means, {\it
i.e.}\ that they are cohomologically trivial, is a consequence the fact that it
is possible to renormalise the classical currents (6.3) to give currents that
{\it do} close at the quantum level.

     So far, we have been concerned here only with the question of
matter-dependent anomalies.  This is an issue that does not even arise for
usual formulations of two-dimensional gravity, since the Virasoro symmetry
is usually realised linearly ({\it i.e.}\ the currents are quadratic in
matter fields).  We must still face the analogues of the anomaly that one {\it
does} meet in two-dimensional gravity, namely the universal anomaly that is
removed by choosing a $c=26$ matter realisation in order to cancel against the
$-26$ contribution to the total central charge coming from the gravity
ghosts.  For $W_\infty$ gravity, we face a more serious-looking problem, since
now there will be ghosts associated with the fixing of the gauge symmetry for
each of the gauge-fields $A^i$.  The ghosts for a spin-$s=i+2$ gauge field
contribute  $$
c_{\rm gh}(s)=-12s^2+12s-2\eqno(6.14)
$$
to the ghostly central charge.  Summing over all $s\ge2$ would seem to imply
that the total ghostly central charge is $c_{\rm gh}({\rm tot})=-\infty$.
At best, an infinity of matter fields seem to be needed, and even then, the
process of cancelling the universal anomaly could be a delicate one.  There
is, however, a different approach that one could take, and that is to adopt
the regularisation scheme that was described in section (5.2).  If this is
done, then we find that the regularised ghostly central charge is $c_{\rm
gh}=2$.

     Assuming for now that the $c_{\rm gh}({\rm tot})=2$ result is to be taken
seriously for $W_\infty$, it follows that the cancellation of the universal
anomalies will occur provided that the matter realisation of $W_\infty$ has
central charge $c_{\rm mat}=-2$.  Remarkably, this is precisely what we have
for our single-scalar realisation!  One can easily check from (6.9) that the
background-charge term is precisely such as to give $c=-2$.  Thus, in a
regularised sense at least, the $W_\infty$ gravity theory that we have
constructed is free of all anomalies [33].  This includes not only the spin-2
anomaly, of the form $(c_{\rm gh}({\rm tot})+c_{\rm mat})\int k_0\partial^3
A_0$, but also the higher-spin anomalies, of the form $C_i\int k_i
\partial^{i+3} A_i$.  For the same reasons as discussed in section (5.2), all
of the coefficients $C_i$ will vanish simultaneously, provided that $c_{\rm
mat}$ is equal to $-2$.

     For now, the possible cancellation of the regularised universal anomlies
should perhaps be viewed as an amusing observation that may ultimately turn out
to have some deeper underlying explanation.  Perhaps the more important lesson
to be derived from looking at the quantisation of classical $w_\infty$ gravity
is that the key requirements are that one should be able to renormalise the
classical currents so that they close on an algebra at the full quantum level.
This ensures the absence of matter-dependent anomalies.   Furthermore, if the
central charge for the matter currents is chosen to cancel that from the ghosts
for the gauge fields, then the universal anomalies will cancel also.  These
{\it desiderata} can be summarised succinctly in one equation:  we require that
the BRST operator $Q$ should be nilpotent.

\bigskip
\noindent{\it 6.2 $W_3$ gravity}
\bigskip

    The philosophy for quantising classical $w_3$ gravity is essentially the
same as that of the previous section.  The starting point is the classical
matter Lagrangian [34]
$$
L=\ft12\bar\partial \varphi_i \,\partial\varphi_i -h T- B W,\eqno(6.15)
$$
where $\varphi_i$ denotes a set of $n$ real matter fields; $h$ and $B$ are
spin-2 and spin-3 gauge fields; and the spin-2 and spin-3 matter currents $T$
and $W$ are given by
$$
\eqalign{
T&=\ft12 \partial\varphi_i \,\partial\varphi_i,\cr
W&=\ft13 d_{ijk}\partial\varphi_i  \,\partial\varphi_j\,\partial\varphi_k.\cr}
\eqno(6.16)
$$
The quantity $d_{ijk}$ is a totally-symmetric constant tensor that satisfies
$$
d_{(ij}{}^k d_{\ell m)k}=\mu \delta_{(ij}\delta_{\ell m)}.\eqno(6.17)
$$
    At the classical level (single contractions), the currents generate what
we may call the $w_3$ algebra,
$$
\eqalign{
\hbar^{-1}T(z)T(w)&\sim {\partial T\over z-w} +{2T\over (z-w)^2},\cr
\hbar^{-1}T(z)W(w)&\sim {\partial W\over z-w} +{3W\over (z-w)^2} ,\cr
\hbar^{-1}W(z)W(w)&\sim {\partial \Lambda \over z-w} + {2\Lambda \over
(z-w)^2}, \cr}\eqno(6.18)
$$
where $\Lambda$ is the composite current
$$
\Lambda=2\mu (T T).\eqno(6.19)
$$
This $w_3$ algebra is a classical limit of the full $W_3$ algebra given in
(2.1$a$-$c$).

     Various possible solutions for the tensor $d_{ijk}$, satisfying (6.17),
have been found [35].  They fall into two categories.  The first consists of
solutions for an arbitrary number of scalars $n$, with the components of the
(totally symmetric) tensor $d_{ijk}$ given by
$$
d_{111}=n,\qquad\qquad d_{1ab} =-n \delta_{ab};\quad
(a=2,\ldots,n).\eqno(6.20)
$$
This satisfies (6.17) with $\mu=n^2$.  The second category of solution relies
upon the abnormalities and perversities of the Jordan algebras.  There are four
such solutions, with $n=5$, $8$, $14$ and $26$ scalars, corresponding to
invariant tensors of Jordan algebras over the reals, complex numbers,
quaternions and octonions respectively [35].  For the complex case, with
$n=8$,  the $d_{ijk}$ tensor coincides with the symmetric $d_{ijk}$ tensor of
$SU(3)$.

    As in the case of the currents (6.3) that generate the $w_\infty$
contraction of $W_\infty$ classically, so also here the currents (6.16)
generate the $w_3$ contraction of $W_3$ classically.  At the full quantum level
of multiple contractions in the operator-product expansion, one finds that the
classical $w_3$ currents (6.16) fail to close on any algebra.  This is the
signal for potential trouble with matter-dependent anomalies upon quantisation
of the theory.  The remedy is again to look for quantum renormalisations of the
currents (6.16) to give currents that do generate an algebra at the quantum
level.  Modulo the freedom to make field redefinitions, the answer, as in the
$w_\infty$ case, is unique.  In this case, it turns out that the resulting
algebra on which the renormalised currents close is $W_3$, given by (2.1).

    The possible renormalisations of the currents (6.16) can be parametrised by
$$
\eqalign{
T&=\ft12\partial\varphi^i \partial\varphi^i +\sqrt\hbar \alpha_i
\partial^2\varphi^i,\cr
W&=\ft13 d_{ijk}\partial\varphi^i \partial\varphi^j \partial
\varphi^k +\sqrt\hbar e_{ij} \partial\varphi^i \partial^2\varphi^j +\hbar f_i
\partial^3 \varphi^i.\cr}\eqno(6.21)
$$
The requirement that the currents generate the $W_3$ algebra gives a set of
conditions on the coefficients $d_{ijk}$, $\alpha_i$, $e_{ij}$ and $f_i$ that
may be found in [35].  The upshot is that the general family of classical
currents, with $d_{ijk}$ given by (6.20) for the case of $n$ scalars, can be
successfully renormalised to give currents that close at the quantum level, on
the full $W_3$ algebra [35].  We shall give the form of the renormalised
currents below.  For the four exceptional cases based on the Jordan algebras,
however,  it appears that it is not possible to renormalise the currents with
$\hbar$-dependent corrections so as to achieve closure [35,36].  This includes
the special 8-scalar realisation based on the totally-symmetric $d_{ijk}$
tensor
of $SU(3)$. There is thus no sense in which the currents in these exceptional
cases could be said to be $W_3$ currents. Although a complete proof of the
impossibility of constructing $W_3$ realisations based on the
quaternionic and octonionic Jordan algebras is still lacking, it has been
proven that in all four exceptional cases there can be no background charges,
and hence the central charges will be 5, 8, 14 and 26 respectively [35].
Consequently, even if the classical currents could be successfully
renormalised, to avoid matter-dependent anomalies, the quantum theory of $w_3$
gravity for any of the four exceptional cases would definitely suffer from
universal anomalies, since anomaly freedom requires that $c=100$.  On the other
hand, for the general family of $n$-scalar realisations with $d_{ijk}$ given by
(6.20), as we shall see below, all anomalies can be cancelled [37].

      Having established that the matter currents for the $n$-scalar
realisations (6.16), (6.20) can be renormalised to give currents that close, on
the $W_3$ algebra, we are now able to proceed to the next stage in the
quantisation procedure, which is to ensure that the matter currents yield an
anomaly-free realisation of the $W_3$ algebra.  This means, as we saw in
section (5.3), that the matter central charge should be $c=100$. A 2-scalar
realisation, with background charge that can be tuned to give, in particular,
$c=100$, was obtained in [5] by making use of the quantum Miura
transformation.  The most general known realisations in terms of scalar fields
are the $n$-scalar realisations found in [35], which correspond to the
renormalisations (6.21) of the classical currents (6.16) with $d_{ijk}$ given
by (6.20).  At $c=100$, these take the form
$$
T=T_X+ \ft12 (\partial\varphi_1)^2+\ft12(\partial\varphi_2)^2 +
\sqrt\hbar \big( \alpha_1\partial^2\varphi_1 +\alpha_2\partial^2\varphi_2
 \big)\eqno(6.22a)
$$
$$
\eqalignno{
W&={2\over \sqrt{261}}\Big\{  \ft13(\partial\varphi_1)^3
-\partial\varphi_1 (\partial\varphi_2)^2 +\sqrt\hbar \big(\alpha_1
\partial\varphi_1  \partial^2\varphi_1 -2\alpha_2 \partial\varphi_1
\partial^2\varphi_2 - \alpha_1 \partial\varphi_2 \partial^2 \varphi_2 \big)\cr
&\qquad\qquad +\hbar\big( \ft13\alpha_1^2\partial^3\varphi_1
-\alpha_1\alpha_2 \partial^3\varphi_2 \big)
-2\partial\varphi_1\, T_X -\alpha_1 \sqrt\hbar\,  \partial T_X\Big\},&(6.22b)
\cr}
$$
where $T_X$ is a stress tensor for $D=n-2$ scalar fields without background
charges,
$$
T=\ft12 \sum_{\mu=1}^D \partial X^\mu\, \partial X^\mu,\eqno(6.23)
$$
and the background charges $\alpha_1$ and $\alpha_2$ for $\varphi_1$ and
$\varphi_2$ are given by
$$
\eqalign{
\alpha_1^2&=-\ft{49}{8}\cr
\alpha_2^2&=\ft1{12}(D-\ft{49}2).\cr}\eqno(6.24)
$$
These conditions on the background charges ensure that the matter central
charge satisfies
$$
c=D+(1-12\alpha_1^2)+(1-12\alpha_2^2)=100.\eqno(6.25)
$$
Note that no matter how many scalar fields one chooses, including $n=100$, it
is necessary to have background charges in order to achieve $c=100$.

     Now that we have obtained a nilpotent BRST operator, and appropriate
matter realisations of the $W_3$ algebra, it is completely straightforward to
write down a Lagrangian for anomaly-free $W_3$ gravity.  We shall not give the
detailed result here; it may be found in [37].  Here, we just remark that it is
obtained from the general BRST prescription:
$$
\lagr=\ft12\bar\partial\varphi^i\partial\varphi^i
 -h T -B W
+\delta\Big( b(h-h_{\rm back}) +\beta(B-B_{\rm back})\Big),
\eqno(6.26)
$$
where $\delta$ denotes the BRST variation, which can be deduced from the BRST
operator (5.24), and $h_{\rm back}$ and $B_{\rm back}$ denote background
gauge-fixed values for the spin-2 and spin-3 gauge fields $h$ and $B$.  Thus
one has [37]
$$
\eqalign{
\lagr&=\ft12\bar\partial\varphi^i\partial\varphi^i
-b{\bar\partial c} -\beta {\bar
\partial}\gamma \cr &\quad +\pi_b(h-h_{\rm back}) + \pi_\beta(B-B_{\rm back})
- h(T+T_{\rm gh}) - B(W+W_{\rm gh}),\cr
}\eqno(6.27)
$$
where $\delta b=\pi_b$ and $\delta \beta=\pi_\beta$.  As in section 2, the
$\hbar$-independent terms in (6.27) (with matter currents $T$ and $W$ given by
(6.22$a$,$b$), and ghost currents $T_{\rm gh}$ and $W_{\rm gh}$ given by
(5.26$a$,$b$)) represent the classical Lagrangian (plus ghost Lagrangian), and
the $\hbar$-dependent terms correspond to counterterms necessary for the
explicit removal of the potential anomalies.

\bigskip
\noindent{\it 6.3  Discussion}
\bigskip

     In sections (6.1) and (6.2), we have reviewed some of the aspects of the
quantisation of $W_\infty$ and $W_3$ gravities.  Our discussion has been
concerned entirely with {\it chiral} $W$ gravity, in the sense that we have
considered gaugings only of a single (left-moving) copy of the algebra.  As
remarked at the beginning of section (6.1), it is completely straighforward to
extend all of the discussions in this paper to the non-chiral case by
exploiting the ingenious trick, introduced in [32], of using additional,
auxiliary, fields.  Thus, for example, for $W_3$ gravity we introduce
auxiliary fields $J^i$ and $\tJ^i$, and write the classical Lagrangian as
[32,38]
$$
\eqalign{
\lagr&=-\ft12\bar\partial\varphi^i\partial\varphi^i-J^i\tJ^i
+\tJ^i\partial\varphi^i+J^i\bar\partial\varphi^i\cr
&\quad -\ft12hJ^iJ^i-\ft13Bd_{ijk}J^iJ^jJ^k-\ft12\th\tJ^i\tJ^i-\ft13\tB
d_{ijk}\tJ^i\tJ^j\tJ^k,\cr}\eqno(6.28)
$$
where the tilded variables refer to a second (right-moving) copy of the gauge
algebra. The equations of motion for the auxiliary fields are
$$\eqalign{J^i&=\partial\varphi^i
-\th \tJ^i -\tB d_{ijk}\tJ^j \tJ^k,\cr
\tJ^i&=\bar\partial\varphi^i
-h J^i -B d_{ijk} J^j  J^k,\cr}\eqno(6.29)
$$
which can be recursively solved to give
$J^i$ and $\tJ^i$ as non-polynomial expressions in $\varphi^i$ and the gauge
fields.  Upon quantisation, one finds that the auxiliary fields have the
propagators [38]
$$
\eqalign{
J^i(z)J^j(w)&\sim {\hbar\delta^{ij}\over (z-w)^2},\cr
\tJ^i(\bar z)\tJ^j(\bar w)&\sim {\hbar\delta^{ij}\over (\bar z-\bar w)^2},\cr
J^i(z)\tJ^j(\bar w)&\sim 0.\cr}\eqno(6.30)
$$
Thus the whole problem has been cloven into separate left-moving and
right-moving sectors [38].  The left-moving matter and ghost currents are now
constructed, at the full quantum level, by replacing $\partial\varphi^i$ in
(6.16), {\it etc.}, by $J^i$.  Similarly, one uses $\tJ^i$ in the construction
of analogous right-moving currents.  Full details may be found in [38].

     An obvious application for anomaly-free $W_3$ gravity is in the
construction of the $W_3$ extension of string theory, {\it i.e.}\ $W_3$
strings. The idea is that the equations of motion for the spin-2 and spin-3
gauge fields impose the vanishing of the spin-2 and spin-3 currents.  At the
quantum level, these conditions can be interpreted, as in ordinary string
theory, as operator constraints on physical states.  By interpreting the
scalar fields in the matter realisations (6.22$a$,$b$) as spacetime
coordinates, one arrives at a first-quantised  description of $W_3$-string
excitations in an $n$-dimensional spacetime.  Because of the necessity for
background charges one does not have the full $SO(1,n-1)$ Lorentz group
acting, but only $SO(1,n-3)$.  The issues arising in the analysis of the
spectrum of $W_3$ strings are quite involved. Preliminary discussions were
given in [38], and a more extensive analysis is contained in [39]. Many
aspects have also been discussed in [40].  Here, we just summarise a couple
of the main results.

     One can see from the form of the realisations (6.22$a$,$b$) that the
scalar $\varphi_1$ is on a very special footing.  In fact all the remaining
scalars ($\varphi_2$ and $X^\mu$) appear only {\it via} their stress tensor.
(In the case of $\varphi_2$, it has a background-charge contribution in the
stress tensor.)  Thus, in some sense the scalar $\varphi_1$ is the only one
which is intrinsically ``non-stringy'' in nature.  The spin-3 current $W$ can
be written as a sum of terms that involve only $\varphi_1$, plus terms
involving the total stress tensor $T$.  It is then rather easy to see at the
classical level that having imposed the $T=0$ constraint, the $W$ constraint
reduces to the statement that $\varphi_1={\rm constant}$ [38].  At the
first-quantised level, this becomes the statement that physical states cannot
involve any $\alpha_{-n}$ creation operators in the $\varphi_1$ direction
[39]. The conclusion from this is that the sole effect of the $W$ constraint
is to ``freeze out'' the $\varphi_1$ coordinate.  The generalisation to $W_3$
strings  introduces the new feature of the non-stringy coordinate $\varphi_1$,
but this is then removed again by the new $W$ constraint.  In fact one is
essentially left with a theory looking very like ordinary string theory,
except that the ``effective'' central charge is $25\ft12$ instead of $26$,
and the ``effective'' $L_0$ intercept is either $1$ or $\ft{15}{16}$.  It
seems that $W_3$ string theory is closely related to ordinary critical string
theory, with the $c=26$ central charge achieved by taking spacetime (with
background charges) to contribute $c=25\ft12$, and adjoining a $c=\ft12$
minimal model [40].  In general, for a $W_N$ string, it seems that it will
be closely related to an ordinary string with a $c=1-\ft6{N(N+1)}$ minimal
model.

     It would be interesting to see what
happens for a supersymmetric extension of the $W_3$ algebra.  This is
currently under investigation [41,42].

     There are other aspects of the quantisation of $W$-gravity theories that
we have not touched on here.  In particular, there is the very interesting
problem of constructing the $W_3$ analogue of the Polyakov induced action of
two-dimensional gravity [43].  To do this, one wants to choose a matter
realisation with non-critical value for the central charge, so that the
universal anomaly ``brings to life'' the Liouville field (and its higher-spin
analogues).  Considerable progress in this area has been made [44]. There seems
to be a certain sense in which there is really no such thing as a
``non-critical'' theory, since the Liouville fields will always come to the
rescue and make up the deficit in the central charge.  It would be interesting
to see whether there is ultimately a convergence of the critical and
non-critical approaches.  Similar issues have also been considered for
$W_\infty$ and $W_{1+\infty}$ gravity.  In particular, it has been shown that
the hidden $SL(2,R)$ Kac-Moody symmetry of light-cone two-dimensional gravity
[45] generalises to $SL(\infty,R)$ for the $W_\infty$ case, and $GL(\infty,R)$
for  the $W_{1+\infty}$ case [46].

\bigskip
\centerline{\bf ACKNOWLEDGMENTS}
\bigskip

     I am very grateful to all my collaborators, and others, for discussions.
These include:  Eric Bergshoeff, Paul Howe, Keke Li, Hong Lu, Larry Romans,
Stany Schrans, Ergin Sezgin, Shawn Shen, Kelly Stelle, Xujing Wang, Kaiwen Xu
and Kajia Yuan.

\vfill\eject

\singlespace
\centerline{\bf REFERENCES}
\frenchspacing
\bigskip

\item{[1]}P. Ginsparg, ``Applied Conformal Field Theory,'' in {\it Les
Houches Lectures}, 1988.

\item{[2]}M.E. Peskin, ``Introduction to String and Superstring Theory,''
Lectures presented at the 1986 Theoretical Advanced Study Institute, Santa
Cruz, 1986.

\item{[3]}T. Banks, ``Lectures on Conformal Field Thoery,'' in {\it Santa Fe
1987 Proceedings}.

\item{[4]}A.B. Zamolodchikov, {\sl Teo. Mat. Fiz.} {\bf 65} (1985) 347.

\item{[5]}V.A. Fateev and A. Zamolodchikov, {\sl Nucl. Phys.}  {\bf
B280} (1987) 644.

\item{[6]}V.A.\ Fateev and S.\ Lukyanov,  {\sl Int. J. Mod. Phys.} {\bf
A3} (1988) 507.

\item{[7]}C.N. Pope, L.J. Romans and X. Shen, {\sl Phys. Lett.} {\bf 236B}
(1990) 173.

\item{[8]}C.N. Pope, L.J. Romans and X. Shen, {\sl Nucl. Phys.} {\bf B339}
(1990) 191.

\item{[9]}C.N. Pope, L.J. Romans and X. Shen, {\sl Phys. Lett.} {\bf
254B} (1991) 401.

\item{[10]}F. Bais, P. Bouwknegt, M. Surridge and K. Schoutens, {\sl
Nucl. Phys.}\ {\bf B304} (1988) 348; 371.

\item{[11]}H.G. Kausch and G.M.T. Watts, {\sl Nucl. Phys.} {\bf B354} (1991)
740.

\item{[12]}R. Blumehagen {\it et al}., {\sl Nucl. Phys.} {\bf B361}( 1991)
255.

\item{[13]}S. Schrans, ``Extensions of Conformal Invariance in
Two-dimensional Quantum Field Theory,'' PhD thesis, Leuven, 1991.

\item{[14]}I. Bakas, {\sl Phys. Lett.} {\bf 228B} (1989) 57.

\item{[15]}L.C. Biedenharn and J.D. Louck, ``The Racah-Wigner Algebra in
Quantum Theory," Addison-Wesley (1981).

\item{[16]}E. Bergshoeff, M. Blencowe and K.S. Stelle, {\sl Comm. Math.
Phys.} {\bf 128} (1990) 213.

\item{[17]}C.N. Pope, L.J. Romans and X. Shen, {\sl Phys. Lett.} {\bf 245B}
(1990) 72.

\item{[18]}E. Witten, {\sl Comm. Math. Phys.} {\bf 113} (1988) 529.

\item{[19]}E. Bergshoeff, C.N. Pope, L.J. Romans, E.
 Sezgin and X. Shen, {\sl Phys. Lett.} {\bf 245B} (1990) 447.

\item{[20]}D. Depiruex, {\sl Phys. Lett.} {\bf 252B} (1990) 586.

\item{[21]}E. Bergshoeff, B. de Wit and M. Vasiliev, {\sl Phys. Lett.} {\bf
256B} (1991) 199;\nl ``The structure of the super-$W_\infty(\lambda)$
algebra,'' preprint, CERN TH-6021-91.

\item{[22]}I. Bakas and E. Kiritsis, {\sl Nucl. Phys.} {\bf B343} (1990) 185.

\item{[23]}M.\ Fukuma, H.\ Kawai and R.\ Nakayama, ``Infinite-dimensional
Grassmannian structure of two-dimensional quantum gravity,'' preprint, UT-572.

\item{[24]}M. Kato and K. Ogawa, {\sl Nucl.  Phys.} {\bf B212} (1983)
443.

\item{[25]}C.N. Pope, L.J. Romans and X. Shen, {\sl Phys. Lett.} {\bf
254B} (1991) 401.

\item{[26]}K. Yamagishi, {\sl Phys. Lett.} {\bf 266B} (1991) 370.

\item{[27]}C.N. Pope, L.J. Romans, E. Sezgin and X. Shen, {\sl Comm. Math.
Phys.} {\bf 140} (1991) 149.

\item{[28]}M.J. Bowick and S.G. Rajeev, {\sl Phys. Rev. Lett.} {\bf 58} (1987)
535; {\sl Nucl. Phys.} {\bf B293} (1987) 348.

\item{[29]}J. Thierry-Mieg, {\sl Phys. Lett.}  {\bf 197} (1987) 368.

\item{[30]}K. Schoutens, A. Sevrin and P. van Nieuwenhuizen, {\sl Comm.
Math. Phys.} {\bf 124} (1989) 87.

\item{[31]}E. Bergshoeff, C.N. Pope, L.J. Romans, E. Sezgin, X. Shen and
K.S. Stelle, {\sl Phys. Lett.} {\bf 243B} (1990) 350.

\item{[32]}K. Schoutens, A. Sevrin and P. van Nieuwenhuizen, {\sl Phys.
Lett.} {\bf 243B} (1990) 245.

\item{[33]}E. Bergshoeff, P.S. Howe, C.N. Pope, E. Sezgin, X. Shen and
K.S. Stelle, {\sl Nucl. Phys.} {\bf B363} (1991) 163.

\item{[34]}C.M. Hull, {\sl Phys. Lett.} {\bf 240B} (1989) 110.

\item{[35]}L.J. Romans, {\sl Nucl.  Phys.} {\bf B352} (1991) 829.

\item{[36]}N.\ Mohammedi, ``On gauging and realising classical and quantum
$W_3$ symmetry,''  ICTP preprint, IC-91-49.

\item{[37]}C.N. Pope, L.J. Romans and K.S. Stelle, {\sl Phys. Lett.} {\bf
268B} (1991) 167.

\item{[38]}C.N. Pope, L.J. Romans and K.S. Stelle, {\sl Phys. Lett.} {\bf
269B} (1991) 287.

\item{[39]}C.N. Pope, L.J. Romans E. Sezgin and K.S. Stelle, ``The $W_3$
String Spectrum,'' preprint, CTP TAMU-68/91, to appear in Phys. Lett. B.

\item{[40]}S.R.\ Das, A.\ Dhar and S.K.\ Rama, ``Physical properties of $W$
gravities and $W$ strings,'' preprint, TIFR/TH/91-11;\nl ``Physical states
and scaling properties of $W$ gravities and $W$ strings,''\nl
preprint, TIFR/TH/91-20.

\item{[41]}H. Lu, C.N. Pope, X.J. Wang and K.W. Xu, ``Anomaly freedom and
realisations for super-$W_3$ strings,'' preprint, CTP TAMU-85/91.

\item{[42]}H. Lu, C.N. Pope, X.J. Wang and K.W. Xu, in preparation.

\item{[43]}A.M. Polyakov, {\sl Mod. Phys.\ Lett.} {\bf A2} (1987) 893.

\item{[44]}H. Ooguri, K. Schoutens, A. Sevrin and P. van Nieuwenhuizen,
``The induced action of $W_3$ gravity,'' preprint, ITP-SB-91-16.

\item{[45]}V.G. Knizhnik, A.M. Polyakov and A.B. Zamolodchikov, {\sl
Mod. Phys. Lett.} {\bf A3} (1988) 819.

\item{[46]}C.N. Pope, X. Shen, K.W. Xu and K. Yuan, ``$SL(\infty,R)$
symmetry of quantum $W_\infty$ gravity,'' preprint, CTP TAMU-37/91, and
Imperial/TP/90-91/29.

\end